\documentclass[12pt, a4paper] {article}

\usepackage{amsfonts}
\usepackage{verbatim}

\begin{document}
\begin{flushright}
arXiv: 0905.0934 [hep-th]\\ CAS-PHYS-BHU/Preprint
\end{flushright}

\vskip 0.5cm

\begin{center}

{\bf {\Large On free 4D Abelian 2-form and anomalous 2D Abelian 1-form gauge 
theories}}

\vskip 1cm

{\bf{ S. Gupta$^{(a)}$, R. Kumar$^{(a)}$, R. P. Malik$^{(a,b)}$}}\\
{\it $^{(a)}$Physics Department, Centre of Advanced Studies,}\\
{\it Banaras Hindu University, Varanasi - 221 005, India}\\

\vskip 0.1cm

{\bf and}\\

\vskip 0.1cm

{\it $^{(b)}$DST Centre for Interdisciplinary Mathematical Sciences,}\\
{\it Faculty of Science, Banaras Hindu University, Varanasi - 221 005, India}\\
{\small e-mails: guptasaurabh4u@gmail.com, raviphynuc@gmail.com, malik@bhu.ac.in}             

\end{center}

\vskip 1 cm

\noindent

\noindent
{\bf Abstract:} We demonstrate a few striking similarities and some glaring differences 
between (i) the free four (3 + 1)-dimensional (4D) Abelian 2-form gauge theory,
and (ii) the anomalous two (1 + 1)-dimensional (2D) Abelian 1-form gauge theory, 
within the framework of Becchi-Rouet-Stora-Tyutin (BRST) formalism. We demonstrate 
that the Lagrangian densities of the above two theories transform in a similar fashion
under a set of symmetry transformations even though they 
are endowed with a drastically different variety of constraint structures. Taking the help of 
our understanding of the 4D Abelian 2-form gauge theory, we prove that the gauge 
invariant version of the anomalous 2D Abelian 1-form gauge theory is a {\it new} 
field-theoretic model for the Hodge theory where all the de Rham cohomological 
operators of differential geometry find their physical realizations in the language 
of proper symmetry transformations. The corresponding conserved charges obey an 
algebra that is reminiscent of the algebra of the cohomological operators. We
briefly comment on the consistency
of the 2D anomalous 1-form gauge theory in the language of 
restrictions on the {\it harmonic} state of the (anti-) BRST and (anti-) co-BRST
invariant version of the above 2D theory.\\ 



\newpage

\noindent
{\bf{\large 1 Introduction}}\\

\noindent
The Becchi-Rouet-Stora-Tyutin (BRST) formalism is one of the most elegant and 
intuitive methods that is required for the covariant canonical quantization of any 
arbitrary $p$-form ($p = 1,2,3...$) gauge and/or reparameterization invariant theories 
that are endowed with the first-class constraints in the language of Dirac's prescription 
for the classification scheme [1,2]. In this formalism, the unitarity and ``quantum" 
gauge (i.e. BRST) invariance are respected together [3-5] at any arbitrary order of
perturbative computation for a given physical process that is allowed by the 
above type of theories.

In recent years, the Abelian 2-form $ ( B^{(2)} =  \frac {1}{2} (dx^\mu \wedge dx^\nu)
 B_{\mu \nu})$ gauge theory with antisymmetric $ ( B_{\mu \nu} = 
- B_{\nu \mu}) $ potential $ B_{\mu \nu} $ [6,7] has become quite popular because 
of its relevance in the context of (super) gravity and (super) string 
theories [8-10]. Its existence is crucial for the emergence of noncommutativity
in the realm of string theory [11]. Furthermore, it has been shown that this gauge
theory provides a field-theoretic model for the quasi-topological field theory [12]
and the Hodge theory [13-15]. This gauge theory, endowed with the first class 
constraints [16], has also been discussed in the framework of BRST formalism [17-19]. 
The (anti-) BRST symmetry transformations for this theory, however, have been shown to be 
anticommutating  only up to a $U(1)$ vector gauge transformation (see, e.g. [13]).

We have applied the superfield formalism to the 4D  Abelian 2-form (and 3-form) gauge 
theories in our recent endeavor  [20]. One of the key outcomes of our work in [20] 
is that the nilpotent (anti-) BRST symmetry transformations must be {\it{absolutely}}
anticommuting because they are identified with the translational generators along the 
Grassmannian directions of the (4,2)-dimensional supermanifold on which the 4D theory 
is generalized. This aspect has been obtained because of the existence 
of a Curci-Ferrari (CF) type restriction [21] that emerges due to the application of 
superfield approach to the above 2-form gauge theory. As is well-known, the 
original CF condition [21] was invoked to ensure the anticommutativity of
(anti-) BRST symmetry transformations in the context of {\it non-Abelian} 1-form gauge theory in 4D.

We have been able to capture the CF type restriction in the Lagrangian 
formulation and have shown explicitly the existence of the absolutely anticommuting
(anti-) BRST  transformations for the free 4D Abelian 2-form 
gauge theory [22]. Added to this, we have been able to demonstrate the
connection of the CF type restriction to the concepts of gerbs that have 
become fairly relevant in the context of string theories. In our present
investigation, we shall exploit the mathematical beauty of the coupled Lagrangian 
densities [22] and show their relevance in the context of anomalous 2D
Abelian 1-form gauge theory [23-25] for the specific set of symmetry considerations.

It is interesting to point out that we have proposed, in our earlier works [26,14],
an alternative set of Lagrangian densities for the 4D Abelian 2-form gauge
theories which are more economical than the ones proposed  in [22]. 
However, in our present endeavor, it is the Lagrangian densities of [22] that 
have the features that are reminiscent of the specific properties associated with the 
anomalous 2D Abelian gauge theory [25]. To be precise, as it turns out, under
the ordinary $U(1)$ gauge transformations, the Lagrangian density of the bosonized 
version of the 2D anomalous gauge theory transforms to a total spactime derivative plus a 
term which is nothing but the off-shoot of 
the Euler-Lagrange equations of motion, derived from the very
same Lagrangian density. Exactly the same feature appears for 
the basic Lagrangian densities of the 4D Abelian 2-form gauge theory [22] under 
a specific set of symmetry transformations within the BRST approach (see, Appendices B and C below).

The central theme of our present investigation is to establish an underlying
mathematical similarity between the free 4D Abelian 2-form gauge theory and the
anomalous 2D Abelian 1-form gauge theory. For this purpose, we focus on the 
(anti-) BRST invariant Lagrangian densities, proposed in our earlier work [22],
where a Lagrange multiplier vector field has been incorporated  
to obtain, in a single step, the CF type restriction that is required 
for the absolute anticommutativity of the nilpotent (anti-) BRST symmetries. 
In addition, a consistent transformation on
this multiplier field ensures the {\it perfect} symmetry invariance 
of the coupled Lagrangian densities of the theory. The other set of Lagrangian 
densities, that are proposed in our earlier works [14,26], play no meaningful role
in our present endeavor.

We demonstrate that, under the nilpotent (anti-) BRST and (anti-) co-BRST symmetry
transformations, the basic Lagrangian densities of the 4D Abelian 2-form gauge 
theory transform to a total spacetime derivative plus a term  which turns out 
to be the equation of motion for the same Lagrangian densities. This feature is 
exactly same as the one we encounter in the case of the bosonized version of the 
2D anomalous Abelian 1-form gauge theory. To be precise, the Lagrangian density 
of the latter theory transforms exactly as the former theory under the
(dual-) gauge, (anti-) BRST and (anti-) co-BRST symmetry transformations (cf. Sec. 3 below).
Furthermore, we demonstrate that {\it {perfectly}} (anti-) BRST 
and (anti-) co-BRST invariant version of the free 4D Abelian 2-form and gauge invariant version 
of the 2D anomalous Abelian 1-form theories are the cute field theoretical
models for the Hodge theory where symmetry considerations play an important role. 
We compare and contrast these theories in Sec. 4 and pin-point explicitly the high
degree of similarities and decisive features of differences between them.  

In our present endeavor, for the first time, we demonstrate the existence of the 
dual-gauge and dual-BRST symmetry transformations for the gauge invariant version [29] 
of the anomalous 2D Abelian gauge theory. This 2D gauge invariant and bosonized 
version of the chiral Schwinger model (CSM), to the best of our knowledge, is proven 
to be a field-theoretic model for Hodge theory for the first time. 
The physical state of the theory
is chosen to be the most symmetric (i.e. harmonic) state of any arbitrarily Hodge decomposed state (of
the total quantum Hilbert space of states). The physicality criteria on this state with 
the BRST and co-BRST charges demonstrate that the anomalous 2D Abelian gauge theory is a consistent 
theory because the physical (harmonic) state is annihilated by the 
individual terms (and their time derivatives) of the expression that appears in the
anomalous behavior [23-25] of the 2D theory (see, Sec. 3 below for details).

Our present investigation is essential on the following counts. First and foremost,
it is always very important to explore a web of mathematical and/or theoretical 
relationships between two different and distinct theories. Our present paper does 
provide some mathematical similarities between 4D Abelian 2-form gauge theory 
and the anomalous 2D Abelian 1-form gauge theory. Second, we propose a set of 
different looking Lagrangian densities for the Abelian 2-form gauge theory 
where the beauty of the mathematical properties of the (anti-) BRST and 
(anti-) co-BRST symmetries are exploited in an elegant manner. These Lagrangian 
densities are different from our earlier Lagrangian densities [22,26,14]. 
Both the above sets, however, have their own importance and individuality. Third, we provide a 
{\it new} field theoretical model for the Hodge theory in 2D which is inspired by 
our understanding of the 4D Abelian 2-form gauge theory. The new field-theoretic 
model happens to be the gauge invariant version of the anomalous 2D gauge theory. 
Four, the physicality condition on the harmonic state proves the consistency of
the anomalous 2D Abelian theory because the anomaly term and its time derivative
annihilate the physical (harmonic) state.
Finally, we discuss,  the similarities and differences between the above two theories.
These  observations might turn out to be useful in our main goal of studying the higher 
$p$-form $(p \geq 3)$ gauge theories within the framework of BRST formalism.

Our present paper is organized as follows. Our second section is dedicated 
to the description of the symmetry properties of the free 4D Abelian 2-form gauge theory. 
This study, ultimately, enables us to prove that the present theory is a field-theoretic model for
Hodge theory. In Sec. 3, we discuss, in detail, some of the key features 
associated with the gauge invariant version of the anomalous 2D Abelian 1-form 
gauge theory. The subject matter of our Sec. 4 concerns itself with the 
discussion of the striking similarities and glaring differences between the above 
two theories. Finally, in our Sec. 5, we summarize our key results, discuss a bit
about some subtle issues present in our endeavor and 
point out a few future directions for further investigations.

Our Appendix A provides a synopsis of the (dual-) gauge transformations
that exist for the 4D Abelian 2-form gauge theory.
In Appendices B and C, we discuss about the derivation of the coupled
Lagrangian densities of this theory that respect nilpotent 
and absolutely anticommuting (anti-) BRST and (anti-) co-BRST symmetries
{\it together}. \\

\noindent 
{\bf {\large 2 Free 4D Abelian 2-form gauge theory: symmetries}}\\

\noindent
In this section, we first discuss the absolutely anticommuting (anti-) BRST
and (anti-) co-BRST
symmetry transformations in subsection 2.1. Our subsection
2.2 is devoted to the discussion of a bosonic symmetry transformation. In subsection
2.3, we discuss the discrete  and ghost scale symmetry transformations. Finally, our
subsection 2.4 deals with the algebraic structure obeyed by the symmetry operators.\\

\noindent
{\bf \large 2.1 Absolutely anticommuting 
(anti-) BRST and (anti-)

co-BRST symmetries: Lagrangian formulation}\\

\noindent
The coupled Lagrangian densities, that respect the nilpotent and absolutely
anticommuting (anti-) BRST as well as (anti-)
co-BRST symmetry transformations {\it together}, are\footnote{We adopt here the notations such that
Greek indices $\mu, \nu, \kappa....= 0, 1, 2, 3$ stand for the spacetime directions
of the 4D flat Minkowski manifold with a metric that possesses signatures (+1, -1, -1, -1)
and the 4D Levi-Civita tensor $\varepsilon_{\mu\nu\eta\kappa}$ is taken with convention
$\varepsilon_{0123} = + 1$. We
also follow $B \cdot \bar B = B_\mu \bar B^\mu \equiv B_0 \bar B_0 - B_i \bar B_i$
where Latin indices $i, j, k....= 1, 2, 3$ correspond to the space directions only.} 
(see, Appendices B and C below for more details)
\begin{eqnarray}
{\cal L}_{(B , \cal B)}^{(L, M)} &=& \frac {1}{2} \partial_\mu \phi_2 
\partial^\mu \phi_2- \frac {1}{2} {\cal B^\mu} \varepsilon _{\mu \nu \eta \kappa }
\partial^\nu B^{\eta \kappa}- \frac {1}{2} ({\cal B} \cdot {\cal B} 
+ \bar {\cal B} \cdot \bar {\cal B}) \nonumber\\
&+& B^\mu (\partial^\nu B_{\nu \mu})+ \frac {1}{2} (B \cdot B + \bar B \cdot \bar B)
- \frac {1}{2} \partial_\mu \phi_1 \partial^\mu \phi_1   
+ \partial_\mu \bar \beta \partial^\mu \beta \nonumber\\
&+& (\partial_\mu \bar C_\nu - \partial_\nu \bar C_\mu ) (\partial^\mu C^\nu )
+(\partial \cdot C - \lambda ) \rho + (\partial \cdot \bar C + \rho ) \lambda \nonumber\\ 
&+& L^\mu (B_\mu - \bar B_\mu - \partial_\mu \phi_1) + M^\mu ({\cal B}_\mu - 
\bar {\cal B}_\mu- \partial_\mu \phi_2), 
\end{eqnarray}
\begin{eqnarray}
{\cal L}_{(\bar B , \bar {\cal B})}^{(L, M)} &=& \frac {1}{2} \partial_\mu \phi_2 
\partial^\mu \phi_2- \frac {1}{2} {\bar {\cal B}^\mu} \varepsilon _{\mu \nu \eta \kappa }
\partial^\nu B^{\eta \kappa}- \frac {1}{2} ({\cal B} \cdot {\cal B} 
+ \bar {\cal B} \cdot \bar {\cal B}) \nonumber\\
&+& \bar B^\mu (\partial^\nu B_{\nu \mu})+ \frac {1}{2} (B \cdot B + \bar B \cdot \bar B)
- \frac {1}{2} \partial_\mu \phi_1 \partial^\mu \phi_1   
+ \partial_\mu \bar \beta \partial^\mu \beta \nonumber\\
&+& (\partial_\mu \bar C_\nu - \partial_\nu \bar C_\mu ) (\partial^\mu C^\nu )
+(\partial \cdot C - \lambda ) \rho + (\partial \cdot \bar C + \rho ) \lambda \nonumber\\
&+& L^\mu (B_\mu - \bar B_\mu - \partial_\mu \phi_1) + M^\mu ({\cal B}_\mu - 
\bar {\cal B}_\mu- \partial_\mu \phi_2),
\end{eqnarray} 
where $L_\mu$ and $ M_\mu$ are the Lorentz vector Lagrange multiplier fields 
and ${\cal B}_\mu, \bar {\cal B}_\mu, B_\mu, \bar B_\mu $ are 
the Nakanishi-Lautrup type auxiliary Lorentz vector fields. The above vector fields
are bosonic in nature. The Lorentz vector fermionic 
($ C_\mu ^2 = \bar C_\mu ^2 = 0, \; C_\mu C_\nu + C_\nu C_\mu = 0, 
\; C_\mu \bar C_\nu + \bar C_\nu C_\mu = 0, $ etc.) (anti-) ghost  fields 
$(\bar C_\mu) C_\mu $ as well as the Lorentz scalar bosonic (anti-) ghost 
fields $(\bar \beta) \beta $ are needed for the validity of unitarity 
(at any arbitrary order of the perturbative calculations). The auxiliary ghost 
fields $\rho$ and $\lambda $ are fermionic  (i.e. $\; \rho \lambda 
+ \lambda \rho = 0, \; \rho^2 = \lambda^2 = 0 $) in  nature 
and massless $(\Box \phi_1 = 0, \Box \phi_2 = 0)$ scalar fields $\phi_1 $ 
and $ \phi_2 $ are required for the stage-one reducibility in the theory
(see, e.g. [14] for more discussions). It is to be noted that the 
totally antisymmetric curvature
tensor $H_{\mu\nu\eta} = \partial_\mu B_{\nu\eta} + \partial_\nu B_{\eta\mu}
+ \partial_\eta B_{\mu\nu}$ is hidden in the above Lagrangian density
in a subtle manner through $\varepsilon_{\mu\nu\eta\kappa} \partial^\nu B^{\eta\kappa}
= (1/3) \varepsilon_{\mu\nu\eta\kappa} H^{\nu\eta\kappa}$.

These 
Lagrangian densities, respecting maximum number of symmetries,  are completely {\it new} for 
the 4D Abelian 2-form gauge theories which have totally different appearance than 
the ones proposed in [14,20,22,26]. It can be checked that the Lagrangian densities 
(1) and (2) respect the following off-shell nilpotent $(s_{(a)b}^2 = 0) $ and 
absolutely anticommuting $(s_b s_{ab} + s_{ab} s_b = 0)$ (anti-) BRST 
transformations $(s_{(a)b})$ 
\begin{eqnarray}
&& s_b B_{\mu \nu} = -(\partial_\mu C_\nu - \partial_\nu C_\mu ), \quad 
s_b C_\mu = - \partial_\mu \beta, \quad s_b \bar C_\mu = -B_\mu, \nonumber\\ 
&& s_b \phi_1 = \lambda , \qquad s_b \bar \beta = - \rho, \qquad 
s_b \bar B_\mu = - \partial_\mu \lambda, \quad 
s_b L_\mu = - \partial_\mu \lambda, \nonumber\\
&& s_b \big [\rho, \lambda, \beta, \phi_2, B_\mu, {\cal B}_\mu, \bar {\cal B_\mu}, 
M_\mu, H_{\mu\mu\kappa}\big ] = 0, 
\end{eqnarray} 
\begin{eqnarray}
&& s_{ab} B_{\mu \nu} = -(\partial_\mu \bar C_\nu - \partial_\nu \bar C_\mu ), \quad
s_{ab} \bar C_\mu = - \partial_\mu \bar \beta , \quad 
s_{ab} C_\mu =  + \bar B_\mu , \nonumber\\
&& s_{ab} \phi_1 = \rho , \qquad s_{ab} \beta = - \lambda , \qquad   
s_{ab} B_\mu = \partial_\mu \rho , \quad s_{ab} L_\mu = - \partial_\mu \rho, \nonumber\\ 
&& s_{ab} \big [\rho , \lambda, \bar \beta, \phi_2, {\cal B}_\mu, \bar {\cal B}_\mu, 
\bar B_\mu , M_\mu , H_{\mu\mu\kappa}\big ] = 0, 
\end{eqnarray} 
because of the fact that the following explicit transformations are valid:
\begin{eqnarray}
s_b {\cal L}_{(B , \cal B)}^{(L, M) } = - \partial_\mu \Big[(\partial^\mu C^\nu 
- \partial^\nu C^\mu )B_\nu + \lambda B^\mu 
+ \rho \partial^\mu \beta \Big ],
\end{eqnarray}
\begin{eqnarray}
s_{ab}{\cal L}_{(\bar B , \bar {\cal B})}^{(L, M)} = -\partial_\mu 
\Big[(\partial^\mu \bar C^\nu - \partial^\nu \bar C^\mu ) \bar B_\nu  
- \rho \bar B^\mu + \lambda \partial^\mu \bar \beta \Big ].
\end{eqnarray}
Under the BRST and anti-BRST symmetry transformations,
the curvature tensor $H_{\mu\nu\kappa}$ and the massless scalar field $ \phi_2  $ 
remain invariant. Thus, the total kinetic term, owing its origin to the
exterior derivative, remains invariant under the (anti-) BRST symmetry transformations.
It is, therefore, concluded that the (anti-) BRST symmetries are the analogue of the
exterior derivative. For more discussion on this issue, we refer
the reader to our earlier work [14]. It is to be remarked that the absolute
anticommutativity of the (anti-) BRST symmetry transformations imply that
only one of them would be really the analogue of the exterior derivative
(see, equations (28), (69) below).

In a similar fashion, it can be seen that the following off-shell nilpotent
$ (s_{a(d)}^2 = 0) $ and absolutely anticommuting 
$ (s_d s_{ad} + s_{ad} s_d = 0)$ (anti-) co-BRST symmetry transformations $ (s_{(a)d}) $
\begin{eqnarray}
&& s_d B_{\mu \nu} = - \varepsilon_{\mu \nu \eta \kappa} \partial^\eta \bar C^\kappa , 
\; s_d \bar C_\mu = - \partial_\mu \bar \beta , \; s_d C_\mu = - {\cal B}_\mu , \nonumber\\
&&  s_d \phi_2 = - \rho , \;  s_d \beta = - \lambda , 
s_d \bar {\cal B}_\mu =  \partial_\mu \rho , \; s_d M_\mu = - \partial_\mu \rho, \nonumber\\ 
&& s_d \big [\rho , \lambda, \bar \beta, \phi_1, 
{\cal B}_\mu, B_\mu, \bar B_\mu, (\partial^\nu B_{\nu \mu}), L_\mu \big ] = 0,
\end{eqnarray}
\begin{eqnarray}
&& s_{ad} B_{\mu \nu} = - \varepsilon_{\mu \nu \eta \kappa} \partial^\eta C^\kappa, \quad  
s_{ad} C_\mu = \partial_\mu \beta, \quad  s_{ad} \bar C_\mu = \bar {\cal B}_\mu , \nonumber\\
&& s_{ad} \phi_2 = - \lambda , s_{ad} \bar \beta = \rho, \; 
s_{ad} {\cal B}_\mu = - \partial_\mu \lambda , \; 
s_{ad} M_\mu = - \partial_\mu \lambda, \nonumber\\ 
&& s_{ad} \big [\rho , \lambda, \beta, \phi_1, 
\bar {\cal B}_\mu, \bar B_\mu,B_\mu, (\partial^\nu B_{\nu \mu}), L_\mu \big ] = 0, 
\end{eqnarray}
leave the Lagrangian densities quasi-invariant because 
\begin{eqnarray}
s_d{\cal L}_{(B,  {\cal B})}^{(L, M)} = \partial_\mu 
\Big[(\partial^\mu \bar C^\nu - \partial^\nu \bar C^\mu ) {\cal B}_\nu  
- \rho {\cal B}^\mu - \lambda \partial^\mu \bar \beta \Big ],
\end{eqnarray}
\begin{eqnarray}
s_{ad}{\cal L}_{(\bar B , \bar {\cal B})}^{(L, M)} = \partial_\mu 
\Big[(\partial^\mu  C^\nu - \partial^\nu  C^\mu ) \bar {\cal B}_\nu  
+ \lambda \bar {\cal B}^\mu + \rho \partial^\mu  \beta \Big ].
\end{eqnarray}
It is evident that the gauge-fixing term ($\partial^\nu B_{\nu \mu}),$ owing its 
origin to the co-exterior derivative, and the field $ \phi_1 $ remain invariant
under the nilpotent (anti-) co-BRST symmetry transformations. It can be 
explicitly checked that 
$ \delta B ^{(2)} = - * d * B ^{(2)} \equiv (\partial^\nu B_{\nu\mu}) d x^\mu $ 
where $ \delta = - * d * $ is the co-exterior derivative and ($*$) is the 
Hodge duality operation. In fact, the nomenclature of (anti-) co-BRST symmetry
transformations owes its origin  to the co-exterior derivative (see, e.g. [14]). 
Thus, the (anti-) co-BRST symmetry transformations are the analogue of the co-exterior 
derivative of differential geometry. The absolute anticommutativity of the (anti-) co-BRST
symmetry transformations, however, imply that only one (of these two transformations)
would be identified with the co-exterior derivative (see, equations (28), (69) below).\\

\noindent
{\bf 2.2 Anticommutator of fermionic symmetries: a bosonic symmetry}\\

\noindent
Our present theory is endowed with a set of four fermionic type $ (s_{(a)b}^2 = 0,
s_{(a)d}^2 = 0) $ symmetry transformations $ s_{(a)b} $ and $s_{(a)d} $. 
It can be explicitly checked that the following operator equations are
true, namely; 
\begin{eqnarray}
\{ s_b , s_{ab} \} = 0, \quad  \{ s_b , s_{ad} \} = 0, \quad 
\{ s_d , s_{ab} \} = 0, \quad \{ s_d , s_{ad} \} = 0,
\end{eqnarray}
when they are applied on any arbitrary field of the theory. Furthermore, we have 
to impose the field  equations $ B_\mu - \bar B_\mu 
- \partial_\mu \phi_1 = 0$ and $ {\cal B}_\mu - \bar {\cal B}_\mu - \partial_\mu 
\phi_2 = 0 $ for the validity of (11) which emerge from the Lagrangian densities 
(1) and/or (2) as equations of motion with respect to $L_\mu$ and $M_\mu$.

The operator $ s_\omega = \{ s_b, s_d \} $ is a bosonic  
type symmetry transformation. The following
infinitesimal version of this  transformation 
\begin{eqnarray}
& s_\omega  B_{\mu \nu } = \partial_\mu {\cal B}_\nu - \partial_\nu {\cal B}_\mu 
+ \varepsilon _{\mu \nu \eta \kappa} \partial^\eta B^\kappa, \; \qquad \;
s_\omega C_\mu = \partial_\mu \lambda,  & \nonumber\\
& s_\omega \Big [\rho, \lambda, \phi_1, \phi_2, \beta, \bar\beta, B_\mu, 
\bar B_\mu, {\cal B}_\mu, \bar {\cal B}_\mu, L_\mu, M_\mu \Big ] = 0, \quad
s_\omega \bar C_\mu = \partial_\mu \rho,&
\end{eqnarray}
leaves the Lagrangian density ${\cal L}_{(B ,  {\cal B})}^{(L, M)}$
quasi-invariant because
\begin{eqnarray}
s_\omega {\cal L}_{(B , {\cal B})}^{(L, M)}&=& \partial_\mu 
\Big [{\cal B}^\mu (\partial \cdot B) - B^\mu (\partial \cdot {\cal B}) \nonumber\\
&-& {\cal B}^\kappa \partial^\mu B_\kappa + B^\kappa \partial^\mu {\cal B}_\kappa 
- \lambda \partial^\mu \rho + (\partial^\mu \lambda) \rho \Big ].
\end{eqnarray}
Thus, transformations (12) are the symmetry transformation for our 
present theory because the action corresponding to the Lagrangian density
(1) remains invariant under (12).  These transformations are the analogue 
of the  Laplacian operator and are same as in our earlier work [14].

The anticommutators of the fermionic transformations $ s_{ad} $ 
and $ s_{ab} $ leads to the derivation of an infinitesimal version of a 
bosonic symmetry transformations
$ (s_{\bar \omega}) $ as given below
\begin{eqnarray}
& s_{\bar \omega}  B_{\mu \nu } = - (\partial_\mu \bar {\cal B}_\nu 
- \partial_\nu \bar {\cal B}_\mu 
+ \varepsilon _{\mu \nu \eta \kappa} \partial^\eta \bar B^\kappa ), \qquad 
s_{\bar \omega} C_\mu = - \partial_\mu \lambda, & \nonumber\\
& s_{\bar \omega} \Big [\rho, \lambda, \phi_1, \phi_2, \beta, \bar \beta, B_\mu, 
\bar B_\mu, {\cal B}_\mu, \bar {\cal B}_\mu, L_\mu, M_\mu \Big ] = 0, \quad s_{\bar \omega} \bar C_\mu = - \partial_\mu \rho. &
\end{eqnarray}
It is straightforward to check that $ s_\omega + s_{\bar \omega} = 0 $ on the 
constrained submainfold defined by the field equation $ {\cal B}_\mu - \bar{\cal B}_\mu 
- \partial_\mu \phi_2 = 0 $. Thus, we conclude that the bosonic transformations 
$ s_{\bar \omega}$ are not independent bosonic symmetry transformations 
{\it{vis-{\`a}-vis}} transformations $ s_\omega.$ In other words, we have
the operator relationship $\{ s_b, s_d \} = s_\omega \equiv - \{ s_{ad}, s_{ab} \}$. \\

\noindent
{\bf 2.3 Ghost and discrete symmetries: ramifications}\\

\noindent
In the Lagrangian densities $ {\cal L}_{ (B, {\cal B})}^ {(L, M)} $ and
$ {\cal L}_{ (\bar B, \bar {\cal B})}^ {(L, M)} $, the fields $ \phi_1, \phi_2, 
B_{\mu \nu}, B_\mu, \bar B_\mu , $  ${\cal B}_\mu, \bar {\cal B}_\mu, L_\mu, M_\mu $
have ghost number equal to zero and the (anti-) ghost fields $(\bar \beta) \beta,
(\bar C_\mu) C_\mu $ and $ (\rho ) \lambda $ have ghost number equal to $(\mp 2), 
(\mp 1) $ and $(\mp 1)$, respectively. The ghost part of the Lagrangian densities 
respect the following infinitesimal transformations $(s_g)$ [14]
\begin{eqnarray} 
&& s_g \; \beta = + 2 \;\Sigma \;\beta, \quad s_g \;\bar \beta = - 2 \;\Sigma \;\bar \beta, \quad
s_g \; C_\mu = + \Sigma \;C_\mu, \nonumber\\
&& s_g \;\bar C_\mu = - \Sigma \;\bar C_\mu, \quad s_g \; \rho = - \Sigma \;\rho, \quad  
s_g \; \lambda = + \Sigma \;\lambda,
\end{eqnarray}
where $\Sigma $ is a global scale parameter. In the above, the numerical factors 
$ (\pm 2) $ and $ (\pm 1) $ denote the corresponding ghost number of the ghost field(s). 
It is evident that the fields, having ghost number equal to zero, do not transform at all
under the ghost transformations. Thus, we have the following infinitesimal ghost
transformations $(s_g)$ for all such fields, namely; 
\begin{eqnarray}
s_g \Psi  = 0, \quad \Psi = B_{\mu \nu}, \phi_1, \phi_2, B_\mu, \bar B_\mu, 
{\cal B}_\mu, \bar {\cal B}_\mu, L_\mu, M_\mu.
\end{eqnarray}
We observe that the above Lagrangian densities 
(1) and (2) remain invariant under the transformations
($s_g$) because $ s_g {\cal L}_{(B, {\cal B})}^{(L, M)} = 0 $ and
$ s_g {\cal L}_{(\bar B, \bar {\cal B})}^{(L, M)} = 0 $.

The Lagrangian densities (1) and (2) also respect the following discrete 
symmetry transformations
\begin{eqnarray} 
&& B_{\mu \nu} \rightarrow \mp \frac {i}{2} \varepsilon_{\mu \nu \eta \kappa}
B^{\eta \kappa}, \quad C_\mu \rightarrow \pm i \bar C_\mu, \quad 
\bar C_\mu \rightarrow \pm i C_\mu, \nonumber\\
&& \beta \rightarrow \pm i \bar \beta, \quad \bar \beta \rightarrow \mp i \beta, \quad 
\phi_1 \rightarrow \pm i \phi_2, \quad \phi_2 \rightarrow \mp i \phi_1, \nonumber\\
&& \rho \rightarrow \mp i \lambda, \quad \lambda \rightarrow \mp i \rho, \quad
L_\mu \rightarrow \mp i M_\mu, \quad M_\mu \rightarrow \pm i L_\mu, \nonumber\\
&& B_\mu \rightarrow \pm i {\cal B}_\mu, \quad {\cal B}_\mu \to \mp i B_\mu, \quad
\bar B_\mu \to \pm i \bar {\cal B}_\mu, \quad \bar {\cal B}_\mu \to \mp i \bar B_\mu.  
\end{eqnarray}
The above symmetry transformations play very important role in establishing 
a connection between the symmetries on the one hand and some key concepts of 
the differential geometry on the other. For instance, these discrete symmetry 
transformations are the analogue of the Hodge duality ($*$) operation of differential 
geometry. Under two successive operations of the transformations (17), it is interesting 
to point out that the following relationships are true on the generic fields
of the theory (see, e.g. [27] for details)
\begin{eqnarray}
&& * \;(* \;B) = + B, \qquad B = B_{\mu\nu}, B_\mu, \bar B_\mu, {\cal B}_\mu,
\bar{\cal B}_\mu, \phi_1, \phi_2, L_\mu, M_\mu, \beta, \bar\beta, \nonumber\\
&& * \;(* \;F) = - F, \qquad F = C_\mu, \bar C_\mu, \rho, \lambda,
\end{eqnarray}
where ($*$) corresponds to the discrete symmetry transformations (17).

Thus, we note that the fermionic and bosonic fields of the theory transform 
in a different manner under a couple of successive operations of the discrete 
transformations. This observation plays an important role in the following 
operator relationship (with $ s_{(a)b}^2 =0,  s_{(a)d}^2 = 0 $):
\begin{eqnarray}
s_{(a)d} \; = \; \pm * \; s_{(a)b} \; *, 
\end{eqnarray}
where $ \pm $ signs,  in the above, are decided by such signs in (18) and $ s_{(a)b} $ 
and $ s_{(a)d}$ are the symmetry transformations (3), (4), (7) and (8).
It is evident that the above relationship is the analogue of the relationship
between the cohomological operators $\delta$ and $d$ (i.e. $ \delta = \pm * d * $
with $\delta^2 = d^2 = 0$).\\

\noindent
{\bf 2.4 Conserved currents and charges: Noether theorem}\\

\noindent
According to Noether's theorem, the continuous symmetry transformations 
$ s_{(a)b}, s_{(a)d}, s_g, s_\omega $ would lead to the derivation of the conserved 
currents as   
\begin{eqnarray}
J^\mu _{(b)} &=& (\partial ^\mu \bar C^\nu - \partial ^\nu \bar C^\mu ) \partial_\nu \beta
- \varepsilon^{\mu \nu \eta \kappa} (\partial_\nu C_\eta ) {\cal B}_\kappa 
- \rho \partial^\mu \beta  \nonumber\\
&-& (\partial ^\mu C^\nu - \partial ^\nu C^\mu ) B_\nu  - \lambda \partial^\mu \phi_1 - \lambda L^\mu,
\end{eqnarray}
\begin{eqnarray}
J_{(ab)}^\mu &=& - \rho L^\mu  - \rho \partial^\mu \phi_1  
- \lambda \partial^\mu \bar \beta  - ( \partial^\mu C^\nu - \partial^\nu C^\mu )
(\partial_\nu \bar \beta ) \nonumber\\ 
&-& \varepsilon^{\mu\nu\eta\kappa} (\partial_\nu \bar C_\eta )\bar {\cal B}_\kappa - 
(\partial^\mu \bar C^\nu - \partial^\nu \bar C^\mu ) \bar B_\nu,
\end{eqnarray}
\begin{eqnarray}
J_{(d)}^\mu &=& (\partial^\mu \bar C^\nu - \partial^\nu \bar C^\mu ) {\cal B}_\nu 
- \varepsilon^{\mu \nu \eta \kappa } B_\nu (\partial_\eta \bar C_\kappa) 
- \rho \partial^\mu \phi_2  \nonumber\\
&+& \rho M^\mu - \lambda \partial^\mu \bar \beta 
- (\partial^\mu C^\nu - \partial^\nu C^\mu ) (\partial_\nu \bar 
\beta), 
\end{eqnarray}
\begin{eqnarray}
J_{(ad)}^\mu &=& (\partial^\mu  C^\nu - \partial^\nu  C^\mu ) \bar {\cal B}_\nu 
- \varepsilon^{\mu \nu \eta \kappa } \bar B_\nu (\partial_\eta  C_\kappa) 
- \lambda \partial^\mu \phi_2  \nonumber\\
&+& \lambda M^\mu + \rho \partial^\mu \beta 
- (\partial^\mu \bar C^\nu - \partial^\nu \bar C^\mu ) (\partial_\nu 
\beta),
\end{eqnarray}
\begin{eqnarray}
J^\mu_{(g)} &=& \; 2 \beta \partial^\mu \bar \beta \; - \; 2 \bar \beta \partial^\mu \beta
\; + \;  (\partial^\mu  C^\nu - \partial^\nu C^\mu ) \bar C_\nu \nonumber\\
&+& (\partial^\mu \bar C^\nu - \partial^\nu \bar C^\mu ) C_\nu 
+ C ^\mu \rho - \bar C^\mu \lambda,
\end{eqnarray}
\begin{eqnarray}
J_{(\omega)}^\mu &=& \varepsilon^{\mu \nu \eta \kappa} \Big [(\partial_\nu 
{\cal B}_\eta ){\cal B}_\kappa + (\partial_\nu B_\eta ) B_\kappa \Big ]
+ \partial_\nu \Big [ B^\mu {\cal B}^\nu - {\cal B}^\mu B^\nu \Big ] \nonumber\\
&+& (\partial^\mu \bar C^\nu - \partial^\nu \bar C^\mu ) (\partial_\nu \lambda ) 
- (\partial^\mu C^\nu - \partial^\nu C^\mu ) (\partial_\nu \rho ).
\end{eqnarray} 
It can be checked that $ \partial_\mu J^\mu_{(i)} = 0 $ $(i = b, ab,d, ad , g , \omega)$ 
if we use the equations of motion derived from $ {\cal L}_{(B, 
{\cal B})}^{(L, M)}$ and $ {\cal L}_{(\bar B, \bar {\cal B})}^{(L, M)}$.
For instance, the equations of motion from $ {\cal L}_{(B, {\cal B})}^{(L, M)}$ are 
\begin{eqnarray}
&& \varepsilon^{\mu \nu \eta \kappa} \partial_\eta {\cal B}_\kappa  
+ (\partial^\mu B^\nu - \partial^\nu B^\mu) = 0, \quad \varepsilon^{\mu \nu \eta 
\kappa} \partial_\eta B_\kappa  - (\partial^\mu {\cal B}^\nu - \partial^\nu 
{\cal B}^\mu) = 0, \nonumber\\
&& B_\mu = \frac {1}{2} (\partial_\mu \phi_1 - \partial^\nu B_{\nu \mu}), 
\; \bar B_\mu = - \frac {1}{2} (\partial_\mu \phi_1 + \partial^\nu B_{\nu \mu}), 
\; \partial \cdot \bar B = 0, \; \partial \cdot B = 0, \nonumber\\
&& {\cal B}_\mu = \frac {1}{2} (\partial_\mu \phi_2 - \frac {1}{2}
\varepsilon_{\mu \nu \eta \kappa} \partial^\nu B^{\eta \kappa}),
 \qquad \bar {\cal B}_\mu = - \frac {1}{2} 
(\partial_\mu \phi_2 + \frac {1}{2} \varepsilon_{\mu \nu \eta \kappa} 
\partial^\nu B^{\eta \kappa}), \nonumber\\
&& \partial \cdot {\cal B} = 0, \quad \partial \cdot \bar {\cal B} = 0, 
\quad B_\mu - \bar B_\mu - \partial_\mu \phi_1 = 0, 
\quad {\cal B}_\mu - \bar {\cal B}_\mu - \partial_\mu \phi_2 = 0, \nonumber\\
&& L^\mu = \bar B^\mu, \; M^\mu = - \bar {\cal B}^\mu, \;\partial \cdot L = 0, \;
\partial \cdot M = 0, \; \Box C^\mu = \frac {1}{2} \partial^\mu (\partial 
\cdot C) \equiv \partial^\mu \lambda, \; \nonumber\\
&& \Box \bar C^\mu = \frac {1}{2} \partial^\mu (\partial \cdot \bar C) 
\equiv - \partial^\mu \rho, \quad \rho  = - \frac {1}{2} (\partial \cdot \bar C), 
\quad \lambda  = \frac {1}{2} (\partial \cdot C), \; \Box B_{\mu \nu} = 0, \nonumber\\
&& \Box \beta = 0, \quad \Box \bar \beta = 0, \quad \Box \rho = 0, \quad
\Box \lambda = 0,  \quad \Box \phi_1 = 0, \quad \Box \phi_2 = 0.
\end{eqnarray}
For the Lagrangian density $ {\cal L}_{(\bar B, \bar 
{\cal B})}^{(L, M)}$, however, the equations of motion are same as the above except 
the following additional relationships
\begin{eqnarray}
&& \varepsilon^{\mu \nu \eta \kappa} \partial_\eta \bar {\cal B}_\kappa  
+ (\partial^\mu \bar B^\nu - \partial^\nu \bar B^\mu) = 0, 
\quad L^\mu = - B^\mu, \nonumber\\
&& \varepsilon^{\mu \nu \eta \kappa} \partial_\eta \bar B_\kappa  
- (\partial^\mu \bar {\cal B}^\nu - \partial^\nu \bar {\cal B}^\mu) = 0, 
\quad M^\mu = {\cal B}^\mu.
\end{eqnarray}

It is interesting to point out that the expressions for the conserved 
currents in (20)-(25) look somewhat different from our earlier work [14].
If we exploit the equations of motion $ L^\mu = \bar B^\mu $,  
$ M^\mu = - \bar {\cal B}^\mu $
and the CF type restriction $ B_\mu - \bar B_\mu - \partial_\mu \phi_1 = 0 $, 
however, we find that the expression for $ J^\mu_{(b)} $ and $ J^\mu_{(d)} $ become 
exactly same as in [14]. Similar is the situation  with $ J^\mu_{(ab)} $ and 
$ J^\mu_{(ad)} $ if we use $ L^\mu = - B^\mu $,  $ M^\mu = + {\cal B}^\mu $ and 
$ {\cal B}_\mu - \bar {\cal B}_\mu - \partial_\mu \phi_2 = 0 $. Thus, we conclude 
that the charges (derived from these conserved currents) would be same as in [14]
and their algebraic structure  would be exactly identical to the ones obtained in 
[14]. Thus, the Lagrangian densities (1) and (2) provide a field theoretic model
for the Hodge theory because the following operator algebra is satisfied, namely;
\begin{eqnarray}
&& s_{(a)b}^2 = 0, \quad  s_{(a)d}^2 = 0, \quad   \{ s_b, s_{ad} \} = \{ s_d, s_{ab} \} = 0, \nonumber\\
&& s_\omega = \{ s_b, s_d \}  \equiv - \{ s_{ab}, s_{ad} \}, \quad 
[s_\omega, s_r] = 0, \quad (r = b, ab, d, ad, g), \nonumber\\
&& [s_g, s_b] = + s_b, \; [s_g, s_d] = - s_d, \;  
[s_g, s_{ad}] = + s_{ad},\;  [s_g, s_{ab}] = - s_{ab}.
\end{eqnarray}
The above operator algebra is analogous to the algebra obeyed by the de Rham
cohomological operators of differential geometry\footnote{
It is worthwhile to mention that on a compact manifold without a boundary, 
there are three cohomological operators $d, \delta, \Delta$ of differential 
geometry. These are christened as the exterior derivative, co-exterior 
derivative and Laplacian operator, respectively. They follow the algebra 
$d^2 = \delta^2 = 0, \Delta = (d + \delta)^2, [\Delta, d ] = 0, 
[\Delta, \delta] = 0$ where $ \delta = - * d *  $ on a 4D spacetime manifold. 
The ($*$), in the above, corresponds to  the Hodge duality operation.} [30-32].

We have, ultimately,  the following interpretations for our 
continuous and discrete symmetry transformations 

(i) only one of the nilpotent and absolutely anticommuting
(anti-) BRST symmetry transformations is the analogue
of the nilpotent ($d^2 = 0$) exterior derivative $d$ of differential geometry,

(ii) only one of the nilpotent and absolutely
anticommuting (anti-) co-BRST symmetry transformations are the 
analogue of the nilpotent ($\delta^2 = 0$) co-exterior derivative
$\delta$ of differential geometry,

(iii) the anticommutator (i.e. $\{s_b, s_d \} \equiv - \{s_{ad}, s_{ab} \}$) 
of the two fermionic $(s_{(a)b}^2 = 0, s_{(a)d}^2 = 0)$ transformations leads 
to the definition of a bosonic symmetry transformation which is the analogue 
of Laplacian operator, and

(iv) the discrete symmetry transformations (17) and ensuing equation
(18) provide us the analogue of the relationship between co-exterior
derivative $(\delta)$ and exterior derivative $(d)$ (i.e. $\delta = \pm * d * $).

To sum up, we have the following mappings: 
$ (s_b, s_{ad}) \rightarrow d, \;  (s_d, s_{ab}) \rightarrow \delta,\; 
\{ s_b, s_d \} = - \{ s_{ab}, s_{ad} \} \rightarrow \Delta$ (see, also [14] for details).\\

\noindent
{\bf{\large 3 Anomalous 2D Abelian 1-form gauge theory}}\\

\noindent
We discuss here the gauge and dual-gauge transformations and corresponding BRST and dual-BRST
transformations for the Lagrangian density of the anomalous 2D Abelian 1-form theory
and its gauge invariant version.\\

\noindent
{\bf 3.1 Gauge and dual-gauge transformations: a synopsis}\\

\noindent
Let us begin with the following effective Lagrangian density of the bosonized version 
of the anomalous 2D Abelian 1-form gauge theory\footnote{We adopt here the 
convention such that Minkowski metric $(g^{\mu\nu})$ is 
with signature $(+1, -1)$ and the antisymmetric Levi-Civita tensor 
$\varepsilon_{\mu\nu}$ is with $\varepsilon _{01} = +1 = -\varepsilon^{01}$, 
$\varepsilon_{\mu \nu} \varepsilon^ {\mu \lambda } = -\delta^{\lambda}_{\nu }$, 
etc. In 2D spacetime, the field strength tensor $F_{\mu \nu}$ has only electric 
field (E) as its existing component and the mass dimension of $A_\mu $ as well 
as $\phi $ is zero (i.e. $[ A_\mu ] = [\phi] = 0 $) and that of the electric charge e 
is one (i.e. [e] = [M]). Here the Greek indices $\mu, \nu...= 0, 1$ and Latin 
indices $i, j... = 1$. Thus, we have $\Box = \partial_0^2 - \partial_1^2$ and 
$(\partial \cdot A) = \partial_0 A_0 - \partial_1 A_1$.} [23-25] 
\begin{eqnarray}
{\cal L}_{eff} &=& -\frac {1}{4} F^{\mu \nu}F_{\mu \nu} 
+ \frac {1}{2} \partial_\mu \phi \partial^\mu \phi + \frac {1}{2} a e^2 A_\mu A^\mu 
+ e (g^{\mu\nu} - \varepsilon ^{\mu\nu})
\partial_\mu \phi A_\nu, \nonumber\\
& \equiv& \frac {1}{2} E^2 + \frac {1}{2} \partial_\mu \phi \partial^\mu \phi 
+ \frac {1}{2} a e^2 A_\mu A^\mu + e ( g^{\mu\nu} - \varepsilon ^{\mu\nu})
\partial_\mu \phi A_\nu,
\label{p1}
\end{eqnarray}
where the 1-form $(A^{(1)} = d x^\mu A_\mu)$ defines the gauge potential $A_\mu$ and 
the 2-form $dA^{(1)} = F^{(2)} = \frac {1} {2!} (dx^\mu \wedge dx^\nu) F_{\mu \nu}$ 
(with $ d = dx^\mu \partial_\mu$, $d^2 = 0$ as exterior derivative) leads to the 
definition of the curvature tensor $F_{\mu\nu} = \partial_ \mu A_\nu 
- \partial_\nu A_\mu$. Here `a' is the parameter that shows the ambiguity in the 
regularization of the fermion determinant when the fermionic chiral Schwinger model 
(CSM) is bosonized in terms of the scalar field $\phi$ and the derivative on it.

It is straightforward to note that under the following infinitesimal gauge transformations 
(with gauge parameter $\chi (x)$ ) (see, e.g. [25])
\begin{eqnarray}
\delta_g A_\mu &=& - \partial_\mu \chi(x), 
\qquad \delta_g \phi = + e \chi(x), \nonumber\\
\delta_g E &\equiv& \delta_g F_{\mu\nu} = 0, \qquad \delta_g 
(\partial \cdot A) = - \Box\chi, \label{p2}
\end{eqnarray}
the Lagrangian density (29) transforms as 
\begin{eqnarray}
\delta_g {\cal L}_{eff} &=& - \partial_\mu \big [ e^2 \varepsilon^{\mu\nu} \chi A_\nu 
+ e^2 (a-1) A^\mu \chi - e \varepsilon^{\mu\nu} \phi \partial_\nu \chi \big ] \nonumber\\ 
&+& e^2 \chi \big [ ( a-1) (\partial \cdot A) + \varepsilon^{\mu\nu} 
\partial_\mu A_\nu \big ].
\end{eqnarray}
It can be easily seen that the curvature 
$ F_{\mu\nu}$, owing its origin to the exterior derivative $ d = d x^\mu \partial_\mu $
(with $d^2 = 0)$, remains invariant under the gauge transformations (30).
Furthermore, the following Euler-Lagrange equations of motion,
derived from the Lagrangian density (\ref{p1}), namely;
\begin{eqnarray}
\Box \phi + e ( g^{\mu\nu} - \varepsilon^{\mu\nu}) \partial_\mu A_\nu = 0, 
\quad \partial_\mu F^{\mu\nu} + a e^2 A^\nu + e ( g^{\nu\eta} 
+ \varepsilon^{\nu\eta}) \partial_\eta \phi = 0,
\end{eqnarray}
imply the following relationship (for $e \neq 0$)
\begin{eqnarray}
(a - 1)\;\partial_\mu A^\mu + \varepsilon^{\mu\nu} \partial_\mu A_\nu = 0. 
\end{eqnarray}
Thus, it is clear that, even though the 2D CSM is anomalous
(i.e. endowed with the second-class constraints [24]) it respects usual gauge 
symmetry if the equations of motion (32) are imposed. The relationship
in (33) has also been shown to be true by exploiting the Hamiltonian formalism 
where the Hamiltonian density is shown to commute with the second-class constraints 
of the 2D anomalous gauge theory [24]. At the moment, we do not know the key reason(s)
behind the existence of a symmetry like (31) because the theory is endowed with
only second-class constraints and, therefore, there should not exist any gauge
type symmetry.

It is very interesting to check that, under the following dual-gauge transformations 
(with an infinitesimal parameter $ \Sigma (x) $): 
\begin{eqnarray}
\delta_{dg} A_\mu = - \varepsilon_{\mu\nu} \partial^\nu \Sigma, 
\quad \delta_{dg} \phi = - e \Sigma, \quad \delta_{dg} E = \Box \Sigma, 
\quad \delta_{dg} (\partial \cdot A) = 0,
\end{eqnarray}
the Lagrangian density (29) transforms as follows
\begin{eqnarray}
\delta_{dg} {\cal L}_{eff} &=& \partial_\mu \Big [ e^2 ( a+1) \varepsilon^{\mu\nu} 
A_\nu \Sigma - e^2 A^\mu \Sigma \nonumber\\ 
&-& e \varepsilon ^{\mu\nu} \phi \partial_\nu \Sigma 
+ E \partial^\mu \Sigma - (\partial^\mu E) \Sigma \Big ] \nonumber\\
&+& e^2 \Sigma \Big [ \frac {\Box E}{e^2} +( \partial \cdot A) 
- (a+1) \varepsilon^{\mu\nu} \partial_\mu A_\nu \Big ].
\end{eqnarray}
We christen these transformations as the dual-gauge transformations because 
it is the gauge-fixing term $\partial_\mu A^\mu \equiv (\partial \cdot A)$, owing 
its origin to the dual-exterior derivative, that remains invariant under (34).
In explicit terms, it can be checked that 
$ \delta A^{(1)} = - * d * ( d x^\mu A_\mu ) = (\partial \cdot A) $. 
The equations of motion (32) imply that the following 
relationship is true, namely;
\begin{eqnarray}
\frac {\Box E}{e^2} + ( \partial \cdot A) - (a + 1) \varepsilon^{\mu\nu}
\partial_\mu A_\nu  = 0,
\end{eqnarray}
where $\Box = \partial_0^2 - \partial_1^2$ is the d'Alembertian operator in 2D. 
Thus, we 
note that the Lagrangian density (29) of the 2D anomalous gauge theory respects

(i) the infinitesimal local gauge symmetry transformations (30), and 

(ii) the infinitesimal dual-gauge symmetry transformations (34),

\noindent
 if we impose the equations of motion (32) (and their off-shoots (33), (36)). 
This observation is {\it exactly} same as the ones, we have
encountered, in the context of the Abelian 2-form gauge theory (see, Appendix A).

Furthermore, it should be noted that we have taken the limit $a < < 1$
so that $1 /(a - 1) \sim - (1 + a)$. Throughout the whole body of our text, we shall
stick to this assumption (i.e. $a < < 1$). It is elementary, then, to check that, in
this limit, we have $(\partial \cdot A) - 
(a + 1) \varepsilon^{\mu\nu} \partial_\mu A_\nu = 0$ 
that emerges from (33) and, furthermore,
we also have $\Box E = 0$ which is valid only for $a  < < 1$.
More discussions on the choice of this region of parameter space
is given in Sec. 5.

Before we wrap up this subsection, we would like to mention, in passing,
that the (dual-) gauge transformations for the 4D Abelian 2-form gauge 
theory has been discussed in [28] where we have obtained a specific set 
of restrictions on the infinitesimal local (dual-) gauge parameters for 
the (dual-) gauge invariance in the theory. These restrictions, however, 
can be converted into the product of the above local parameters and
equations of motion corresponding to the specific fields of the theory.
Thus, we claim that there is one-to-one correspondence between the above 
mentioned theories as far as the symmetry properties are concerned (see, Appendix A).

The similarities between the two theories motivate us to look for the existence of the
(anti-) BRST and (anti-) co-BRST symmetries for the 2D theory. This is what precisely
we do in our forthcoming subsections.\\

\noindent
{\bf 3.2 BRST and anti-BRST transformations: gauge invariant CSM}\\

\noindent
Corresponding to the gauge transformations (30), we have the off-shell
nilpotent ($ s_{(a)b} ^ 2 = 0$) and absolutely anticommuting 
$(s_b s_{ab} + s_{ab} s_b = 0 )$ (anti-) BRST symmetry transformations $ s_{(a)b} $
\begin{eqnarray}
&& s_b A_\mu = - \partial_\mu C, \quad s_b C = 0, \quad s_b \bar C = i b, 
\quad s_b b = 0, \quad s_b\phi = e C,\nonumber\\
&& s_b E = s_b F_{\mu\nu} = 0, \quad s_b (\partial\cdot A) = -  \Box C, 
\quad s_b \theta = - e^2 C, \nonumber\\          
&& s_{ab} A_\mu = - \partial_\mu \bar C, \quad s_{ab} \bar C = 0, 
\quad s_{ab} C = - i b, \quad s_{ab} b = 0, \quad s_{ab}\phi = e \bar C,\nonumber\\
&& s_{ab} E = s_{ab} F_{\mu\nu} = 0, \quad s_{ab} (\partial\cdot A) = - 
\Box \bar C, \quad s_{ab} \theta = - e^2 \bar C, 
\end{eqnarray}
under which, the following Lagrangian density (with an additional field $\theta$)
\begin{eqnarray}
{\cal L}_{(b)} &=& \frac {1}{2} E^2 + \frac {1}{2} \partial_\mu \phi
\partial^\mu \phi + \frac {1}{2} a e^2 A_\mu A^\mu + e (g^{\mu\nu} 
- \varepsilon ^{\mu\nu}) \partial_\mu \phi A_\nu \nonumber\\
&+& \theta\; \Big [(a - 1) (\partial \cdot A) + \varepsilon ^{\mu\nu}
\partial_\mu A_\nu \Big ]\; + \;\frac {(a - 1)} {2 e^2} \partial_\mu \theta 
\partial^\mu \theta \nonumber\\
&+& b (\partial \cdot A) + \frac {1} {2} b^2 
+ i \partial_\mu \bar C \partial^\mu C,
\end{eqnarray} 
remains quasi-invariant because it changes to a total spacetime derivative as 
\begin{eqnarray}
s_b {\cal L}_{(b)} &=& \partial_\mu \Big [ (1 - a) e^2 C A^\mu - e^2 \varepsilon ^{\mu\nu} 
C A_\nu + e \varepsilon^{\mu\nu} \phi \partial_\nu C \nonumber\\
&+& (1 - a) \theta \partial^\mu C - b \partial ^\mu C \Big ], \label{sg}
\end{eqnarray}
\begin{eqnarray} 
s_{ab} {\cal L}_{(b)} &=& \partial_\mu \Big [ (1 - a) e^2 \bar C A^\mu 
- e^2 \varepsilon ^{\mu\nu} \bar C A_\nu
+ e \varepsilon^{\mu\nu} \phi \partial_\nu \bar C \nonumber\\
&+& (1 - a) \theta \partial^\mu \bar C  - b \partial ^\mu \bar C \Big ].
\end{eqnarray}
It will be noted that the constrained relationship (33), which was derived in two 
steps from the Lagrangian density (29), is now derived in one step because the 
equation of motion with respect to $\theta$, namely;
\begin{eqnarray}
\frac{(a -1)}{e^2} \; \Box \theta = (a -1)\; (\partial \cdot A) 
+  \;\varepsilon^{\mu\nu}\partial_\mu A_\nu, 
\end{eqnarray}
produces it if we set the limit $\theta \rightarrow 0.$ Furthermore, it is worth 
emphasizing that the mass dimensions of $[\theta] = [M], [b] = [M], [C] = 0, 
[\bar C] = 0$, etc.,  ensure the appropriate mass dimension of
the Lagrangian density (38).

In the above, the $b$ field is the Nakanshi-Lautrup auxiliary field and 
$(\bar C)C$ are the fermionic ($C^2 = \bar C^2 = 0, \; C \bar C + \bar C C = 0$) 
(anti-) ghost fields that are needed for the unitarity in the theory. In the Lagrangian density 
(38), it is clear that the gauge-fixing and Faddeev-Popov terms can be expressed as
\begin{eqnarray}
s_b s_{ab} \Bigl ( - \frac{i}{2} A_\mu A^\mu + \frac{1}{2} C \bar C \Bigr )
= b (\partial \cdot A) + \frac{1}{2} b^2 + i \partial_\mu \bar C 
\partial^\mu C. \label{fp} 
\end{eqnarray}
If we do {\it not} incorporate the terms that contain $\theta$ 
fields in (38), then, under the (anti-) BRST transformations, the Lagrangian density would 
transform as 
\begin{eqnarray}
s_b {\cal L}_{(b)}^{(\theta \rightarrow 0)} &=& - \partial_\mu 
\big [ e^2 \varepsilon^{\mu\nu} C A_\nu 
+ e^2 (a-1) A^\mu C - e \varepsilon^{\mu\nu} \phi \partial_\nu C 
+ b \partial^\mu C \big ] \nonumber\\ 
&+& e^2 C \big [ ( a-1) (\partial \cdot A) + 
\varepsilon^{\mu\nu} \partial_\mu A_\nu \big ],\nonumber\\
s_{ab} {\cal L}_{(b)}^{(\theta \rightarrow 0)} &=& - \partial_\mu \big 
[ e^2 \varepsilon^{\mu\nu} \bar C A_\nu 
+ e^2 (a-1) A^\mu \bar C - e \varepsilon^{\mu\nu} \phi \partial_\nu \bar C 
+ b \partial^\mu \bar C \big ] \nonumber\\ 
&+& e^2 \bar C \big [ ( a-1) (\partial \cdot A) + 
\varepsilon^{\mu\nu} \partial_\mu A_\nu \big ].
\end{eqnarray}
This observation is {\it exactly} same as the one we have encountered in the context of the
4D Abelian 2-form gauge theory (Appendix B) where the Lagrangian density transforms to 
a total spacetime derivative plus a term that is found to be zero on-shell.
Thus, to obtain a perfect BRST symmetry transformation, it is essential to add 
these additional terms containing $\theta$.

There are a few points that have to be emphasized at this stage. First, the 
$\theta$ field here is not a Lagrange multiplier field because it possesses a 
kinetic term and, therefore, is a dynamical field. Second, the motivation for 
adding the term $\theta \big [(a - 1) (\partial \cdot A) + \varepsilon ^{\mu\nu} 
\partial_\mu A_\nu \big ]$ has come from the fact that (33) is the equation of 
motion from our starting Lagrangian density (29). We have incorporated this term 
in the Lagrangian density following our understanding of the Abelian 2-form gauge 
theory where we have incorporated the terms  $L^\mu (B_\mu - \bar B_\mu - \partial_\mu \phi_1)$ 
and $M^\mu ({\cal B}_\mu - \bar {\cal B}_\mu - \partial_\mu \phi_2)$  
in the Lagrangian densities 
(cf. (1), (2)). Finally, it can be seen that under the (anti-) BRST symmetry 
transformations, the equation (33) transforms as
\begin{eqnarray}
s_b \Big [(a - 1) (\partial \cdot A) + \varepsilon ^{\mu\nu} \partial_\mu A_\nu \Big ] 
= - (a - 1) \Box C, 
\end{eqnarray}
\begin{eqnarray}
s_{ab} \Big [(a - 1) (\partial \cdot A) + \varepsilon ^{\mu\nu} \partial_\mu A_\nu \Big ] 
= - (a - 1) \Box \bar C. 
\end{eqnarray}
This is why, a kinetic term for the field $\theta$ has to be added in the Lagrangian 
density to achieve the perfect (anti-) BRST symmetries. Thus, 
we conclude that there is a perfect conceptual analogy between our present 2D theory
and the 4D Abelian 2-form gauge theory (see, Sec. 2 and Appendices B, C). In the latter case, 
it was found that the equations of motion $B_\mu - \bar B_\mu - \partial_\mu \phi_1 = 0$
and $ {\cal B}_\mu - \bar {\cal B}_\mu - \partial_\mu \phi_2 = 0$
were invariant under the nilpotent (anti-) BRST and (anti-) co-BRST symmetry transformations
(i.e. $s_{(a)b} [ B_\mu - \bar B_\mu - \partial_\mu \phi_1] = 0$
and $s_{(a)d} [ {\cal B}_\mu - \bar {\cal  B}_\mu - \partial_\mu \phi_2] = 0$). This is why
there was no need to add a kinetic term for the Lagrange multiplier field $L_\mu$ as well
as $M_\mu$.

Before we close up this subsection, it is worth noting that the Lagrangian density 
of [29], that results in from the gauge-invariant generating functional of the 
bosonized version of the CSM in 2D, is same as the one quoted in (38) modulo some 
constant factors and the gauge-fixing and Faddeev-Popov ghost terms. However,
the logic behind the derivation of the Lagrangian density (38) is totally
different and it has emerged out from our understanding  of the derivation of
the Lagrangian densities (1) and (2) in the context of 4D free Abelian 2-form
gauge theory (see, Sec. 2 and Appendices B and C). It is worthwhile to mention that the inclusion of the $\theta$ field
has, in fact, rendered the second-class constraints of the original bosonized version of
the CSM to the first-class constraints [29]. This is why, there is existence of a perfect 
(anti-) BRST symmetry invariance (cf. (37), (39) and (40)) in the theory.\\

\noindent
{\bf 3.3 (Anti-) dual BRST symmetry transformations: a discussion}\\

\noindent
The BRST invariant Lagrangian density $ {\cal L}_b $ is also endowed with the off-shell
nilpotent $ (s_{(a)d} ^ 2 = 0 )$ (anti-) dual BRST symmetry transformations $s_{(a)d}$. For this 
purpose, we linearize the kinetic term ($ - \frac {1}{4} F^{\mu\nu} F_{\mu\nu} = 
\frac {1}{2} E ^2 $) by invoking an auxiliary field $ \bar b $ in the following fashion
\begin{eqnarray}
{\cal L}_{(b, d)} &=& \bar b E - \frac {1}{2} \bar b^2 + \frac {1}{2} \partial_\mu \phi
\partial^\mu \phi + \frac {1}{2} a e^2 A_\mu A^\mu + e (g^{\mu\nu}
- \varepsilon ^{\mu\nu}) \partial_\mu \phi A_\nu \nonumber\\
&+& \theta \; \Big [(a - 1) (\partial \cdot A) + \varepsilon ^{\mu\nu}
\partial_\mu A_\nu \Big ]\; + \;\frac {(a - 1)} {2 e^2} \partial_\mu 
\theta \partial^\mu \theta \nonumber\\
&+& b (\partial \cdot A)  
+ \frac {1}{2} b^2 + i \partial_\mu \bar C \partial^\mu C.
\end{eqnarray}
It can be checked that the following nilpotent ($ s_{(a)d} ^2 = 0 $) and absolutely 
anticommuting ($ s_d s_{ad} + s_{ad} s_d = 0 $) (anti-) dual BRST transformations
$ s_{(a)d} $
\begin{eqnarray}
&& s_d A_\mu = - \varepsilon_{\mu\nu} \partial ^\nu \bar C, \quad 
s_d \bar C = 0, \quad  s_d C = i \bar b, \quad  s_d \bar b = 0, \nonumber\\
&& s_d \phi = - e \bar C, \quad s_d E = \Box \bar C, \quad s_d (\partial \cdot A) = 0, 
\quad s_d b = 0, \nonumber\\
&& s_d \theta = - \frac {e^2 \bar C} { (a - 1)} \cong   e^2 \bar C (1 + a) , 
\end{eqnarray}
\begin{eqnarray}
&& s_{ad} A_\mu = - \varepsilon_{\mu\nu} \partial ^\nu C, \quad 
s_{ad} C = 0, \quad  s_{ad} \bar C = -i \bar b, \quad  s_{ad} \bar b = 0, \nonumber\\ 
&& s_{ad} \phi = - e C, \quad s_{ad} E = \Box C, \quad s_{ad} (\partial \cdot A) = 0, 
\quad s_{ad} b = 0, \nonumber\\
&& s_{ad} \theta = - \frac {e^2 C} { (a - 1)} \cong  e^2 C (1 + a) , 
\end{eqnarray}
leave the Lagrangian density (46) quasi-invariant because 
\begin{eqnarray}
s_d {\cal L}_{(b, d)} &=& \partial_\mu \Big [ \bar b \partial^\mu \bar C 
- e \varepsilon ^{\mu\nu} \phi \partial_\nu \bar C - \theta \partial ^ \mu
\bar C \nonumber\\ &-&  e^2 A^\mu \bar C + (a + 1) e^2 \bar C \varepsilon ^{\mu\nu} A_\nu \Big ], 
\end{eqnarray}
\begin{eqnarray}
s_{ad} {\cal L}_{(b, d)} &=& \partial_\mu \Big [ \bar b \partial^\mu  C 
- e \varepsilon ^{\mu\nu} \phi \partial_\nu C - \theta \partial ^ \mu C \nonumber\\
&-& e^2 A^\mu  C + (a + 1) e^2  C \varepsilon ^{\mu\nu} A_\nu \Big ]. 
\end{eqnarray} 
Thus, the action corresponding to the above Lagrangian density $ {\cal L}_{(b, d)} $
remains invariant under the (anti-) dual-BRST symmetry transformations $ s_{(a)d}$.

It is an interesting point to note that if $\theta$ terms are {\it not} incorporated
in the Lagrangian density ${\cal L}_{(b,d)}$, the latter would transform, under
the (anti-) co-BRST symmetry transformations, as
\begin{eqnarray}
s_{d} {\cal L}^{(\theta \to 0)}_{(b,d)} &=& \partial_\mu \Bigl [ e^2 (a + 1) \varepsilon^{\mu\nu} 
A_\nu \bar C - e^2 A^\mu \bar C 
- e \varepsilon ^{\mu\nu} \phi \partial_\nu \bar C 
+ \bar b \partial^\mu \bar C  \Bigr ] \nonumber\\
&+& e^2 \bar C \Big [ ( \partial \cdot A) 
- (a+1) \varepsilon^{\mu\nu} \partial_\mu A_\nu \Big ], \nonumber\\
s_{ad} {\cal L}^{(\theta \to 0)}_{(b,d)} &=& \partial_\mu \Big [ e^2 ( a+1) \varepsilon^{\mu\nu} 
A_\nu  C - e^2 A^\mu  C 
- e \varepsilon ^{\mu\nu} \phi \partial_\nu  C 
+ \bar b \partial^\mu  C  \Big ] \nonumber\\
&+& e^2  C \Big [ (\partial \cdot A) 
- (a+1) \varepsilon^{\mu\nu} \partial_\mu A_\nu \Big ],
\end{eqnarray}
which are the analogues of (43) where we have taken $(a - 1)^{-1} \sim - (1 + a)$ because
of the fact that $a < <1$. It is worth pointing out that the nature of transformations in 
the above is {\it exactly} same as the one, we have encountered in the context of  
Abelian 2-form gauge theory (see, Appendix C below).

The analogue of the equation (42) can be written in terms of the 
(anti-) dual-BRST symmetry transformations ($s_{(a)d}$) as 
\begin{eqnarray}
s_d s_{ad} \Bigl ( - \frac{i}{2} A_\mu A^\mu + \frac{1}{2} C \bar C \Bigr )
= \bar b \; E - \frac{1}{2} \bar b^2 + i \partial_\mu \bar C 
\partial^\mu C. 
\end{eqnarray}
Thus, we note that the kinetic term for the gauge field $A_\mu$ and the Faddeev-Popov
ghost terms can be written in the exact-form 
with the help of the (anti-) co-BRST symmetry transformations.
In the above form, the (anti-) co-BRST invariance of the Lagrangian density ${\cal L}_{(b,d)}$
becomes quite simple because of the nilpotency (i.e. $s_{(a)d}^2 = 0$), anticommutativity
(i.e. $s_d s_{ad} + s_{ad} s_d = 0$) and the invariance of the gauge-fixing term
(i.e. $s_{a)d} (\partial \cdot A) = 0, s_{(a)d} b = 0$).\\

\noindent
{\bf{3.4 Bosonic symmetry: anticommutator of fermionic symmetries}}\\

\noindent
In the context of the gauge invariant version of the 2D anomalous gauge theory, 
we have established the existence of four nilpotent (fermionic) symmetries 
(i.e. $s_{(a)b}, s_{(a)d} $). The following 
infinitesimal version of the bosonic (i.e. $s_\omega = \{s_b, s_d\}$) 
transformations $s_\omega$
\begin{eqnarray}
&& s_\omega A_\mu = - i (\varepsilon _{\mu\nu} \partial^\nu b 
+ \partial_\mu \bar b), \quad s_\omega \theta = - i e^2 \Big(\frac  {b}{a -1} 
+ \bar b \Big), \quad s_\omega E = i \Box b, \nonumber\\
&&  s_\omega \phi = + i e  (\bar b - b), \quad s_\omega 
(\partial \cdot A) = - i \Box \bar b, \quad
s_\omega \big [C, \bar C, b, \bar b \big ] = 0,
\end{eqnarray}
is the symmetry transformation in the theory because the Lagrangian density (46)
transforms, under the above infinitesimal transformations, as
\begin{eqnarray}
s_\omega {\cal L}_{(b, d)} &=& \partial_\mu X^\mu,\nonumber\\
X^\mu &=& i (\bar b \partial^\mu b - b \partial^\mu \bar b) 
\; - \; i  \; \theta \partial^\mu \Big [b + (a-1) \bar b \Big ] \nonumber\\
&-& i \; e^2  b A^\mu \; - \;  i \; e^2 \varepsilon^{\mu \nu} \Big [\bar b A_\nu 
+ \frac {1}{(a-1)}  b A_\nu \Big ] \nonumber\\
&-& i \; e^2 (a-1) \bar b A^\mu 
\; + \; i \; e  \varepsilon ^{\mu \nu} \phi \partial_\nu (\bar b - b). 
\end{eqnarray}
As a result, the action of the theory remains invariant under (53).

It can be explicitly checked that the anticommutators $ \{s_d, s_{ad} \} = 0 $,
$ \{s_b, s_{ab} \} = 0 $, $ \{s_d, s_{ab} \} = 0 $,
$ \{s_b, s_{ad}\} = 0 $. Thus, the remaining anticommutator $\{s_{ad}, s_{ab} \} 
= s_{\bar \omega} $ produces a bosonic symmetry transformation $ s_{\bar \omega}$
which is {\it not} independent of $s_\omega$.
It can be explicitly checked that 
\begin{eqnarray} 
(s_\omega  +  s_{\bar \omega})\; \Omega = 0,
\end{eqnarray}
where $ \Omega $ is an arbitrary generic field of the theory. This establishes the fact that
$s_{\bar \omega} = - s_\omega$. In other words, we have $s_\omega = \{ s_b, s_d \} 
\equiv - \{ s_{ad}, s_{ab} \}$. \\

\noindent
{\bf { 3.5 Ghost and discrete symmetries: outcomes}}\\

\noindent
The bosonic fields $ A_\mu, \theta, \phi, b, \bar b $ of the theory have 
ghost number equal to zero whereas the fermionic fields $ C$ and $\bar C$
have ghost number equal to $\pm 1$. Thus, we have the following ghost scale
transformations 
\begin{eqnarray}
&& A_\mu \rightarrow A_\mu, \qquad \theta \rightarrow \theta, \qquad \phi \rightarrow  \phi,
\qquad b \rightarrow b, \nonumber\\
&&  \bar b \rightarrow \bar b , \qquad C \rightarrow e^{+\Lambda} C,
\qquad \bar C \rightarrow e^{-\Lambda} \bar C,
\end{eqnarray} 
under which the Lagrangian density ${\cal L}_{(b, d)} $ remains invariant. 
The numbers $\pm 1$ in the exponentials of $C$ and $\bar C$ transformations correspond
to the ghost numbers and $\Lambda$ is a global infinitesimal scale parameter.

It can be, furthermore, checked that under the following discrete symmetry transformations
\begin{eqnarray}
&& a \rightarrow -a, \quad e \rightarrow \pm i e, \quad C \rightarrow \pm i \bar C, \quad
\bar C \rightarrow \pm i C, \nonumber\\
&& b \rightarrow \pm i \bar b, \qquad  \bar b \rightarrow \pm i b, \qquad  
A_\mu \rightarrow  \pm i \varepsilon_{\mu\nu} \; A^\nu, \nonumber\\
&& \phi \rightarrow \phi, \qquad \theta \rightarrow \pm  \frac {i \theta}{(a + 1)} 
\; \cong \; \mp i \theta (a - 1),
\end{eqnarray}
the Lagrangian density ${\cal L}_{(b, d)}$ remains invariant. Equivalently, under 
another discrete symmetry transformations 
\begin{eqnarray}
&& a \rightarrow -a, \quad e \rightarrow \pm i e, \quad C \rightarrow \pm i \bar C, \quad
\bar C \rightarrow \pm i C, \nonumber\\
&& b \rightarrow \pm i \bar b, \quad  \bar b \rightarrow \pm i b, \quad  
A_\mu \rightarrow  A_\mu, \quad \partial_\mu \rightarrow  \pm i \varepsilon_{\mu\nu}
 \partial^\nu, \nonumber\\
&& \phi \rightarrow \phi, \qquad \theta \rightarrow \pm  \frac {i \theta}{(a + 1)} 
\; \cong \; \mp i \theta (a - 1),
\end{eqnarray}
the Lagrangian density ${\cal L}_{(b, d)}$ remains unchanged. The above discrete symmetry 
transformations (57) and (58) play very important role as it can be clearly seen that the 
following relationships are true, namely;
\begin{eqnarray}
s_{(a)d} \Omega = \pm * s_{(a)b} * \Omega, \quad 
\Omega = A_\mu, \phi, b, \bar b,\theta, C, \bar C,
\end{eqnarray}
where $\Omega$ is an arbitrary generic field of the theory and the nilpotent transformations
$s_{(a)b}$ and $s_{(a)d}$ are explicitly illustrated in (37), (47) and (48). 
For the 2D theory, it can be also checked that the reverse relationship 
$s_{(a)b} \Omega = \pm * s_{(a)d} * \Omega$ is also true (see, e.g. [27] for details).

In the 
above, the ($*$) symbol corresponds to the discrete symmetry transformations (57) 
and/or (58) and the signs $(\pm)$ are dictated by such signs that appear in the two successive
operations of ($*$) as given below
\begin{eqnarray}
 * \; [ * \;(\Omega) ]  = \pm \;\Omega, \qquad \Omega = A_\mu, C, \bar C, \phi,
 \theta, b, \bar b, E, (\partial \cdot A).
\end{eqnarray}
It can be explicitly checked that, with ($ \Omega = \Omega_1, \Omega_2$), we have [27]
\begin{eqnarray}
&& * \; [ * \;(\Omega_1) ] = + \;\Omega_1, \qquad \Omega_1 = \phi, \nonumber\\
&& * \; [ * \;(\Omega_2) ] = -\; \Omega_2, \qquad \Omega_2 = A_\mu, b, \bar b, C, 
\bar C, \theta, (\partial \cdot A), E.
\end{eqnarray} 
The above relations are true with respect to the discrete transformations (57).
A bit different relation emerges with the transformations in (58) where we
find $\Omega_1 = \phi, A_\mu$ and $\Omega_2 =  b, \bar b, C, 
\bar C, \theta, (\partial \cdot A), E.$
The relationship in (59) is the analogue of the relationship between the co-exterior
derivative $(\delta)$ and the exterior derivative $ d $ (i.e. $ \delta = \pm * d * $).
Thus, we note that the analogue of the Hodge duality $(*)$ operation is the discrete
symmetry transformations (57) and/or (58) for the 2D Abelian 1-form gauge theory.\\

\noindent
{\bf { 3.6 Conserved charges and algebra: impacts}}\\

\noindent 
The continuous symmetry transformations, according to Noether's theorem, lead
to the conserved currents. These conserved currents, corresponding to the 
symmetry transformations $s_{a(b)}, s_{a(d)}, s_\omega $ and ghost 
transformations are: 
\begin{eqnarray}
J^\mu_{(b)} &=& a e^2 C A^\mu + F^{\mu\nu} (\partial_\nu C) - b
\partial^\mu C - (a - 1) C (\partial^\mu \theta) \nonumber\\
&+&  e C (\partial^\mu \phi)  - \theta \varepsilon^{\mu\nu} \partial_\nu C
- e \varepsilon^{\mu\nu} \phi (\partial_\nu C), 
\end{eqnarray} 
\begin{eqnarray}
J^\mu_{(d)} &=& e \varepsilon^{\mu\nu}\phi (\partial_\nu \bar C) - a e^2 \bar C
\varepsilon^{\mu\nu} A_\nu - e \bar C (\partial ^\mu \phi)  + \bar b
\partial ^\mu \bar C  \nonumber\\
&-&  \bar C (\partial^\mu \theta) - \theta (a - 1) \varepsilon ^{\mu\nu}
\partial_\nu \bar C - b \; \varepsilon ^{\mu\nu} (\partial_\nu \bar C),
\end{eqnarray}
\begin{eqnarray}
J^\mu_{(\omega)} &=& i e (\bar b - b) \partial^\mu \phi + i e^2 a \bar b A^\mu 
+ i \bar b \; \varepsilon^{\mu\nu} \partial_\nu \bar b 
- i \theta \varepsilon ^{\mu \nu} \partial_\nu \bar b\nonumber\\
&-& i b \varepsilon^{\mu \nu} \partial_\nu b + i e \varepsilon^{\mu \nu} 
\phi \partial_\nu (b - \bar b) - i \Big [b + (a-1) \bar b \Big ]  
\partial^\mu \theta  \nonumber\\ 
&-& i (a - 1) \theta \varepsilon^{\mu \nu} \partial_\nu b 
- i e^2 a b \varepsilon^{\mu \nu} A_\nu,       
\end{eqnarray}
\begin{eqnarray}
J^\mu_{(g)} \; = \; - i (C \partial^\mu \bar C + \bar C \partial^\mu C ),
\end{eqnarray}
where $ J^\mu_{(g)} $ is the ghost Noether current. It will be noted that the expressions
for $ J^\mu_{(ab)} $ and $ J^\mu_{(ad)} $ can be obtained from $ J^\mu_{(b)} $ 
and $ J^\mu_{(d)} $ by the replacements: $ C \rightarrow \bar C $ and 
$ \bar C \rightarrow C $.
The above currents are conserved because it can be checked that 
$ \partial_\mu J^\mu_{(i)} = 0 $ for  $i = b, ab, d, ad, \omega, g $. For this proof,
however, the following equations of motion, emerging from $ {\cal L}_{(b, d)}, $ 
have to be used:
\begin{eqnarray}
&& \varepsilon^{\mu \nu} \partial_\nu \bar b + (a - 1) \partial^\mu \theta 
- \varepsilon^{\mu \nu} \partial_\nu \theta + \partial^\mu b - a e^2 A^\mu
- e ( g^{\mu \nu} + \varepsilon ^{\mu\nu}) \partial_\nu \phi = 0, \nonumber\\ 
&& \Box \theta = \frac{e^2}{(a - 1)} \big[ (a - 1) (\partial \cdot A) 
+ \varepsilon^{\mu\nu} \partial_\mu A_\nu \big ], \quad b = - (\partial \cdot A),
\quad \bar b = E, \nonumber\\
&& \Box \phi + e (g^{\mu\nu} - \varepsilon^{\mu\nu}) \partial_\mu A_\nu = 0, 
\quad \Box C =  \Box \bar C = 0, \quad \Box b =  \Box \bar b = 0 .  
\end{eqnarray}
It is worth mentioning that we have always taken $ a < < 1 $ and consequently
$ 1/(a - 1) \sim - (1 + a)$ has been used throughout the whole body of the text.

Using the equations of motion (66), it is straightforward to obtain the 
expression for the conserved charges $ Q_i = \int J^0_{(i)} dx $ 
$(i = b, ab, d, ad, \omega, g)$ as 
\begin{eqnarray}
&& Q_b =  \int \;dx \; \bigl [ \dot b \;C - b \;\dot C \bigr ], \quad Q_{ab} =  
\int\; dx \;\bigl [ \dot b \;\bar C - b \;\dot {\bar C} \bigr ], \nonumber\\
&& Q_d =  \int\; dx \;\bigl [ \bar b\; \dot {\bar C} - \dot {\bar b}\; \bar C \bigr ], 
\quad Q_{ad} =  \int\; dx \; \bigl [ \bar b \;\dot C - \dot {\bar b } \;C  \bigr ], \nonumber\\
&& Q_{\omega} = - i \int\; dx \; \bigl [ b \;\dot {\bar b}  - \bar b\; \dot b \bigr ], 
\quad Q_g = - i \int\; dx \;\bigl [ C \;\dot {\bar C} + \bar C \;\dot C \bigr ].
\end{eqnarray}
These conserved charges obey an algebra that is reminiscent of the algebra 
of the cohomological operators. These are succinctly expressed as 
\begin{eqnarray}
&& Q^2_{a(b)} = 0, \quad Q^2_{a(d)} = 0,\quad  [Q_{\omega}, Q_r] = 0,
 \quad ( r = b, ab, d, ad, g ),\nonumber\\
&& \{ Q_b, Q_{ab} \} = 0, \quad \{ Q_d, Q_{ad} \} = 0,
\quad  \{ Q_b, Q_{ad} \} = 0,  \nonumber\\
&&  Q_{\omega} = \{ Q_d, Q_{b} \} = - \{ Q_{ad}, Q_{ab} \} ,     
\quad \{ Q_d, Q_{ab} \} = 0,\nonumber\\
&& i [ Q_g, Q_b ] = + Q_b, \quad i [ Q_g, Q_{ab} ] = - Q_{ab}, \nonumber\\
&& i [ Q_g, Q_d ] = - Q_d, \quad i [ Q_g, Q_{ad} ] = + Q_{ad}.
\end{eqnarray} 
Thus, we note that there exists a two-to-one mapping between the conserved charges
on the one hand and the de Rham cohomological operators on the other. This statement can be 
captured in the following set of equations: 
\begin{eqnarray}
&& (Q_b, Q_{ad})\rightarrow d, \qquad (Q_{ab}, Q_d)\rightarrow \delta, \nonumber\\
&& Q_{\omega} = \{ Q_b, Q_d \} = -\{ Q_{ab}, Q_{ad} \} \rightarrow \Delta. 
\end{eqnarray}
It is clear, therefore, that the symmetries and conserved charges are the realizations
of the de Rham cohomological operators. The physical reasons behind the mapping in (69)
are exactly same as the ones we have discussed in the context of the free 4D Abelian 2-form
gauge theory (cf. Sec. 2).

If a state $\mid \psi>_n$, in the quantum Hilbert space, has the ghost number equal to $n$
(i.e. $ i \; Q_g \mid \psi>_n = n \mid \psi>_n $), the following relationships turn out 
to be true if we exploit the algebraic relations (68), namely;
\begin{eqnarray}
&& i Q_g Q_b \; \mid \psi >_n \; = (n + 1) \; Q_b \mid \psi >_n, \nonumber\\ 
&& i Q_g Q_d \; \mid \psi >_n \; = (n - 1) \; Q_d \mid \psi >_n, \nonumber\\
&& i Q_g Q_{ab} \; \mid \psi >_n \; = (n - 1) \; Q_{ab} \mid \psi >_n, \nonumber\\
&& i Q_g Q_{ad} \; \mid \psi >_n \; = (n + 1) \; Q_{ad} \mid \psi >_n, \nonumber\\
&& i Q_g Q_\omega \; \mid \psi >_n \; = n \; Q_\omega \mid \psi >_n.
\end{eqnarray}
Thus, the ghost numbers of the states $ Q_b \mid \psi>_n$,  $Q_d \mid \psi>_n$ and  
$Q_\omega \mid \psi>_n$
are  $(n+1), (n-1)$ and  $n$, respectively. This observation is the analogue of the
basic facts connected with the differential geometry where the degree of an $n$-form ($f_n$)
increases by one, decreases by one and remains intact due to the operations of the
exterior, dual-exterior and  the Laplacian operator, respectively. That is to say,
in the mathematical terms,
we have: $ d f_n \sim f_{n + 1},\; \delta f_n \sim f_{n - 1}$ and $\Delta f_n \sim f_n$,
respectively.

One of the decisive features of the present 2D model of the Hodge theory is that, under 
the discrete symmetry transformations (57) and/or (58), we have the following relationships:
\begin{eqnarray}
&* \; Q_b = + \; Q_d,\quad * \; Q_d = + \; Q_b,\quad * \; 
Q_\omega = + \; Q_\omega, & \nonumber\\
& * \; Q_g = - \; Q_g,\quad * \; Q_{ab} = + \; Q_{ad}, \quad * \; Q_{ad} = + Q_{ab}.& 
\end{eqnarray}
This feature is distinctly different [14] from the 4D Abelian 2-form gauge theory where 
$* \; Q_b = +  Q_d, \; * \; Q_d = - Q_b, \;* \;Q_\omega = -  Q_\omega, * \; Q_{ad} = - Q_{ab},\;
* \; Q_{ab} = + Q_{ad}, \;* \;Q_g = - Q_g $. This difference is connected with the dimensionality of 
the two different theories [27]. It is interesting to point out that the total algebra (68) 
remains invariant under the $(*)$ operation corresponding to the discrete symmetry
transformations listed in (57) and/or (58).\\

\noindent
{\bf { 3.7 Physical state as a harmonic state: consequences }}\\

\noindent 
It is worth pointing out that, consistent with the algebraic structures in (68), (69) 
and (70), one can write an arbitrary state $ \mid  \psi >_n $ (with ghost number $n$)
in the quantum Hilbert space, as the following sum
\begin{eqnarray}
\mid  \psi >_n &=& \mid h >_{(n)} + \; Q_b \mid \chi > _{(n - 1)} 
+ \; Q_d \mid \xi > _{(n + 1)} \nonumber\\
&\equiv & \mid h >_{(n)} + \; Q_{ad} \mid \chi > _{(n - 1)} 
+ \; Q_{ab} \mid \xi > _{(n + 1)},
\end{eqnarray}
where, in the first line, the state $Q_b |\chi>_{(n - 1)}$ is a BRST exact state,
the state $Q_d \mid \xi > _{(n + 1)}$ is the BRST co-exact state and $|h>_{(n)}$
is the harmonic state. A similar kind of statement can be made for the second line.
The above equation is the analogue of the Hodge decomposition  theorem (HDT) [30-32]
which states that any arbitrary $n$-form $f_n$, on a compact manifold without a boundary, 
can be uniquely written as the sum of a harmonic form $h_n$ with 
$(\Delta h_n = 0, \; d h_n = 0, \; \delta h_n = 0)$, an exact form $(d e_{n - 1})$ 
and a co-exact form $(\delta c_{n + 1})$. Mathematically, this statement can be 
expressed as
\begin{eqnarray}
f_n = h_n + d e_{n - 1} + \delta c_{n + 1}.  
\end{eqnarray}
Due to the two-to-one mapping (cf. (69)), however, the HDT can be expressed in two 
different ways in the quantum Hilbert space of states. Taking the help of  mapping
in (69), we have captured this statement in (72).

The most symmetric state, in the quantum Hilbert space of the states, is the harmonic state
$ \mid h >_{(n)}$ in (72) which is annihilated by $Q_{(a)b}, \; Q_{(a)d}$ and
$Q_\omega$. We choose this state as the physical state of the theory (i.e.
$\mid h_n >_{(n)} \;\equiv \; \mid phys >)$. This immediately implies that 
\begin{eqnarray}
Q_\omega \mid phys > \; = 0, \quad Q_{(a)b} \mid phys > \; = 0, \quad 
Q_{(a)d}\mid phys > \; = 0. 
\end{eqnarray}
It will be noted that
all the above restrictions are consistent with one-another.
The latter two relations, in the above, produce the following restrictions on the 
physical state (that are different from the ghost states), namely;
\begin{eqnarray}
b \mid phys > \; = 0, \quad \dot b \mid phys > \; = 0, \quad \bar b \mid phys > \; = 0, 
\quad \dot {\bar b} \mid phys > \; = 0,
\end{eqnarray}
so that the physical state could become symmetric with respect to the nilpotent
and conserved (anti-) BRST and (anti-) co-BRST charges.

It is evident from the equations of motion (66) that the above restriction in (75) 
imply the following restrictions on the physical state
\begin{eqnarray}
&& (\partial \cdot A) \mid phys > \; = 0, \quad \partial_0 (\partial \cdot A) \mid phys > \; = 0, 
\nonumber\\ && E \mid phys > \; = 0, 
\quad \dot E \mid phys > \; = 0.
\end{eqnarray}
Thus, we notice that the anomalous behavior, that appears in the 
r.h.s. of the conservation
law $\partial_\mu J^\mu \sim
\big [(a - 1) (\partial \cdot A) + \varepsilon^{\mu \nu} 
\partial_\mu A_\nu \big ]$ (see, e.g. [24,25]), is trivally zero
because of the physicality condition. Here $J^\mu$
is defined through  $\partial_\nu F^{\nu\mu} = J^\mu$ (which is also equivalent
to $\varepsilon^{\mu\nu} \partial_\nu \bar b = - J^\mu$ because
of the equations of motion and the observation that $F_{\mu\nu} = \varepsilon_{\mu\nu} \bar b$).
The above statement is true because this
conservation law is valid in the quantum Hilbert space as
\begin{eqnarray}
< phys|\; \partial_\mu J^\mu\; |phys> \sim 
< phys|\; \bigl [(a - 1) (\partial \cdot A) + \varepsilon^{\mu\nu} \partial_\mu A_\nu \bigr ]\; |phys>.
\end{eqnarray}
However, as we have seen that $Q_{(a)b}\; |phys> = 0 \;\Rightarrow \;(\partial \cdot A) |phys> \;=\; 0,
\partial_0 (\partial \cdot A) |phys> = 0$ and $Q_{(a)d} |phys> = 0 \Rightarrow E |phys> = 0,
\dot E \;|phys> = 0$, it is clear that the individual terms of the anomalous expression 
(and their time derivatives, too) annihilate the physical state of the theory.

It should be mentioned that the above statements are valid 
in the limit $\theta \rightarrow 0$ which corresponds to the 
true anomalous 2D Abelian 1-form gauge theory. On the face value, the
$\theta$-dependent terms do not appear in the expressions for
$Q_{(a)b}$ and $Q_{(a)d}$. However, they turn up in the
expressions for the time derivatives of $(\partial \cdot A)$ and 
$E = - \varepsilon^{\mu\nu} \partial_\mu A_\nu$ due to the
dynamical equations of motion listed in (66). 
Thus, we conclude that the anomalous 2D Abelian 1-form gauge theory is a consistent theory
because of the physicality conditions on the 
harmonic state with the (anti-) BRST and (anti-) co-BRST charges
(which are conserved and nilpotent of order two).\\

\noindent
{\bf 4. Similarities and differences: a bird's-eye view}\\

\noindent
The two theories, under discussion, are completely different theories in different
dimensions of spacetime. Thus, there are bound to be too many differences. However,
the interesting and amazing aspects of these theories are that they have some
common points of similarities. We point out here some striking similarities 
and key conceptual differences between these theories. In particular, we
concentrate more on the common features of similarity and focus only on the conceptual
issues as far as the differences are concerned.

The first and foremost aspect of similarity is the nature of the 
transformations of the Lagrangian densities 
under the (anti-) BRST and (anti-) co-BRST symmetry transformations. It can be
seen from equations (92), (93), (31) and (43) that, under the nilpotent and absolutely anticommuting
(anti-) BRST symmetry transformations, the Lagrangian densities of the two theories transform
to a total spacetime derivative plus a term that is proportional to one of the equations
of motion (see, Appendix B for (92) and (93)). In exactly similar fashion, 
from equations (105), (106), (35) and (51), it can be
noted that the Lagrangian densities of the two theories behave in exactly the same manner under the nilpotent
and absolutely anticommuting (anti-) co-BRST symmetry transformations
(see, Appendix C for (105) and (106)).

The second feature that draws our attention is that, for the existence of the {\it perfect}
symmetry invariance, we incorporate a couple of terms (e.g. $ L^\mu (B_\mu - \bar B_\mu
- \partial_\mu \phi_1), M^\mu ({\cal B}_\mu - \bar {\cal B}_\mu - \partial_\mu \phi_2)$)
in the Lagrangian  densities, through the Lagrange multiplier fields,
in the case of the free 4D Abelian 2-form gauge theory (cf. (1),(2)). 
The above logic of the 4D Abelian 2-form theory, with a bit of modification, 
also works in the case of anomalous 2D Abelian 1-form gauge theory. In fact, to begin with,
we add a term proportional to the equation of motion 
(i.e. $ \theta [(a- 1) (\partial \cdot A) + \varepsilon^{\mu\nu} \partial_\mu A_\nu]$ with
$\theta$ as a Lagrange multiplier field) in the Lagrangian density of the original theory.
However, this turns out to be insufficient for our purpose. The above features are totally different
from our understanding of the 4D (non-)Abelian 1-form gauge theories 
where there is absolutely {\it no} need of any kind of multiplier fields (see, e.g., [3,4] for details).

Despite our logic being same for both the theories, a bit of difference crops up because of the following
reasons. It is straightforward to note that the field equations $B_\mu - \bar B_\mu - \partial_\mu \phi_1$
and ${\cal B}_\mu - \bar {\cal B}_\mu - \partial_\mu \phi_2$ remain off-shell invariant under the nilpotent
(anti-) BRST and (anti-) co-BRST transformations. The above statement can be mathematically
expressed as 
\begin{eqnarray}
&& s_{(a)b}\; [B_\mu - \bar B_\mu - \partial_\mu \phi_1] = 0, \quad
s_{(a)b}\; [{\cal B}_\mu - \bar {\cal B}_\mu - \partial_\mu \phi_2] = 0, \nonumber\\
&& s_{(a)d}\; [B_\mu - \bar B_\mu - \partial_\mu \phi_1] = 0, \quad
s_{(a)d}\; [{\cal B}_\mu - \bar {\cal B}_\mu - \partial_\mu \phi_2] = 0, 
\end{eqnarray}
where $s_{(a)b}$ and $s_{(a)d}$ are given in (3), (4), (7) and (8). The same does not
hold good with the field equation $ (a - 1) (\partial \cdot A) + \varepsilon^{\mu\nu} \partial_\mu A_\nu$
in the context of anomalous 2D Abelian 1-form theory. This statement, besides (44) and (45), can be
mathematically stated as 
\begin{eqnarray}
&& s_d [(a - 1) (\partial \cdot A) + \varepsilon^{\mu\nu} \partial_\mu A_\nu] = -  \Box \bar C, \nonumber\\
&& s_{ad} [(a - 1) (\partial \cdot A) + \varepsilon^{\mu\nu} \partial_\mu A_\nu] = - \Box C.
\end{eqnarray}
Thus, we note that, in the context of anomalous 2D Abelian theory, the equation of motion
$(a - 1) (\partial \cdot A) + \varepsilon^{\mu\nu} \partial_\mu A_\nu$ remains invariant
under the (anti-) BRST and (anti-) co-BRST symmetry transformations only on the on-shell
(i.e. $\Box C = \Box \bar C = 0 $) for $a \neq 1$ (cf. (44),(45),(79)).

This is the reason that a ``kinetic'' piece,
corresponding to the field $\theta$, has to be incorporated in the Lagrangian density for the
perfect symmetry invariance in the context of anomalous 2D Abelian 1-form gauge theory
(cf. (46)). However, such addition makes the  $\theta$-field as a dynamical (propagating)
field in the theory. It is to be emphasized, at this juncture, that the logic behind the derivation
of the Lagrangian densities (1), (2) and (46) for the 4D Abelian 2-form and anomalous
2D Abelian 1-form theories is the same. Thus, there is a striking similarity between
these two theories. It should be re-emphasized that the above features
 are completely different from our understanding of the 4D
(non-)Abelian 1-form gauge theories where there is no
need to incorporate any kind of CF type restriction {\it explicitly} in the (anti-) BRST
invariant Lagrangian density of the above theories [3,4].

The third point of similarity between the two theories is the observation that the 
modified Lagrangian densities (cf. (1),(2),(46))
of the two theories are endowed with continuous symmetry transformations
and discrete symmetry transformations which render them to be a field theoretic-model
for the Hodge theory. Of course, the original anomalous 2D Abelian 1-form theory is described
by the Lagrangian density that is a limiting case
of the Lagrangian density (46) when $\theta \to 0$. However, the point to be noted is that
both the theories, in some sense, are the {\it modified} versions of the basic theories
(as far as the true philosophy of BRST formalism is concerned).

At the conceptual level, we enumerate here a few key differences between the
two theories. Both the theories are drastically different
in the sense that the free 4D Abelian 2-form gauge theory is endowed with the first-class
constraints (see, e.g. [16]) but the original anomalous 2D Abelian 1-form gauge theory 
possesses only second-class constraints [24] in the language of Dirac's prescription
for classification scheme. Furthermore, they exist in different dimensions of the spacetime.
They are rendered to be the models for the Hodge theory through symmetry considerations. However,
the methods to achieve the {\it perfect} symmetries, in both the theories, are different 
in the sense that the former needs only the Lagrange multipliers fields but the latter
requires the ``kinetic'' piece for the ``Lagrange multiplier'' field as well.

The CF type restrictions, in the context of the 4D 2-form theory, play double roles
because, not only they accomplish the anticommutativity of the (anti-) BRST and (anti-) co-BRST
symmetries, but they also render the theory to possess the maximum number of perfect symmetries. The role
of the equation of motion $(a - 1) (\partial \cdot A) + \varepsilon^{\mu\nu} \partial_\mu A_\nu = 0 $,
on the other hand,
is totally different in the context of anomalous 2D Abelian 1-form theory. Whereas the
CF type restrictions (i.e. $B_\mu - \bar B_\mu - \partial_\mu \phi_1  = 0$ and 
${\cal B}_\mu - \bar {\cal B}_\mu - \partial_\mu \phi_2  = 0$) are derived directly from the 
Lagrangian densities (1) and (2), the condition
$(a - 1) (\partial \cdot A) + \varepsilon^{\mu\nu} \partial_\mu A_\nu = 0$ emerges from
(66) as the limiting case when $\theta \to 0$. Furthermore, in the proof of consistency
of the anomalous 2D Abelian 1-form gauge theory the latter condition plays an important role
(i.e. $\partial_\mu J^\mu \sim (a - 1) (\partial \cdot A) + \varepsilon^{\mu\nu} \partial_\mu A_\nu$).
We have briefly commented about it through the physicality condition
with the conserved and nilpotent (anti-) BRST and (anti-) co-BRST charges
(see, Subsec. 3.7, for details).\\

\noindent
{\bf{5. Summary and discussion}}\\

\noindent
In our present investigation, we have demonstrated the similarity of the coupled Lagrangian 
densities of the free 4D Abelian 2-form gauge theory [22] with the Lagrangian density of the anomalous 2D Abelian gauge theory under a specific set of symmetry transformations. 
To be precise, we have established that the {\it basic} Lagrangian densities\footnote{We call
the Lagrangian densities (91) as {\it basic} because these are similar to the Lagrangian
densities of 4D (non-)Abelian 1-form gauge theories (having no interaction with matter fields) [3,4].
In the latter theories there is {\it no} need of any Lagrange multiplier fields.}
of the 4D Abelian 2-form 
gauge theory transform, under the (anti-) BRST and (anti-) co-BRST symmetry transformations, 
to a total spacetime derivative plus a term that is zero on the equations of motion
that are derived from the coupled Lagrangian densities (cf. (92)-(95),
(105)-(108) in Appends. B and C). This feature is 
{\it exactly} same as the nature of transformations in the context of 
the anomalous 2D Abelian 1-form gauge theory under the (dual-) gauge, (anti-) BRST and (anti-) co-BRST 
symmetry transformations (cf. (31),(43),(51)).

It is to be noted that, only in the context of the 4D Abelian 2-form gauge theory, the extra pieces
(e.g. $L^\mu (B_\mu - \bar B_\mu - \partial_\mu \phi_1)$
and $M^\mu ({\cal B}_\mu - \bar {\cal B}_\mu - \partial_\mu \phi_2)$) have to be incorporated
in the Lagrangian densities (cf. (1),(2)) for the perfect symmetry invariance. On the contrary,
such kind of extra pieces are absolutely {\it not} required for the symmetry invariance in the context
of 4D (non-)Abelian 1-form gauge theories (see, e.g. [3,4]). In fact, the analogue of
(42), in the case of 4D (non-)Abelian 1-form gauge theories, is good enough for the perfect
symmetry invariance. It has been claimed in our earlier work [22] that the CF type
restrictions, in the context of the free 4D Abelian 2-form gauge theory,
have deep connection with the concept of gerbes and they would always
appear in the context of higher-form ($p \geq 2$) gauge theories. In our recent works [20],
the above claim has been shown to be true in the case of the free 4D Abelian 3-form gauge theory.

To obtain the {\it perfect} symmetry invariance,
we have introduced a pair of Lagrange multiplier fields (i.e. $L^\mu$
and $M^\mu$) for the Abelian 2-form gauge theory. A noteworthy point is that, the ``kinetic terms'' for these multiplier fields, are {\it not} required for the perfect symmetry invariance in the theory. 
This is due to the fact
that the CF type restrictions (i.e. equations of motion) remain absolutely invariant under the (anti-) BRST
and (anti-) co-BRST symmetry transformations (cf. (78)). We follow the same trick in the context of anomalous
2D Abelian theory and introduce a Lagrange multiplier field $\theta$. However, the constraint 
conditions (i.e. equations of motion) are {\it not} found to be {\it absolutely} invariant under
the (anti-) BRST as well as (anti-) co-BRST symmetry transformations. Rather, they are
found to be invariant only on the  on-shell conditions $\Box C = \Box \bar C = 0$
(see, Sec. 4 for details).

To circumvent the above difficulty,  we have added a kinetic piece for the field $\theta$  
to obtain the perfect symmetry invariance
in the theory. As a consequence, the $\theta$-field becomes a propagating 
(dynamical) field and it behaves, no longer, as a Lagrange multiplier field.  In fact, it is due to
the presence of the $\theta$-terms that we have been able to show the existence of the (anti-) BRST
and (anti-) co-BRST symmetry transformations for the modified version of the anomalous 2D
Abelian 1-form gauge theory.

The existence of the dual-gauge and (anti-) co-BRST symmetry transformations is a completely
{\it new} result as far as the modified version of the anomalous 2D Abelian 1-form theory is concerned. In fact,
these symmetry transformations enable us to prove that the system,  described by the 
Lagrangian density (46), provides a new field-theoretic model for the Hodge theory. In this
context, it is pertinent to point out that, so far, we have been able to prove the following
field-theoretic models for the Hodge theory: 

(i) the free 2D (non-)Abelian 1-form gauge theories without any interaction
    with matter fields [33-35],
    
(ii) the interacting 2D $U(1)$ Abelian gauge theory with matter fields as Dirac fields [36,37], and

(iii) the free 4D Abelian 2-form gauge theory [13-15].

One of the key assumptions, in our present investigation, has been the choice of the ambiguity
parameter $a$ to be in the region $a < < 1$. In this context, it is to be pointed out that,
in a very recent work [38], it has been demonstrated, with the help of the numerical computation,
that $ a = 1$ is an exceptional point in the theory. We have avoided this point by
our choice $a < < 1$ and have confined ourselves to the region of the parameter space where the
modified version of the anomalous 2D theory respects maximum symmetries which render it
to become  a model for the Hodge theory.

At this juncture, it is worthwhile to mention that we have shown the existence of the dual-BRST
symmetry transformations in the context of the 2D QED with Dirac fields [36,37].
The latter fields undergo an (anti-) BRST version of the  chiral transformations corresponding
to the (anti-) co-BRST symmetry transformations (i.e. $ s_d A_\mu = - \varepsilon_{\mu\nu}
\partial^\nu \bar C, s_{ad} A_\mu =  - \varepsilon_{\mu\nu} \partial^\nu C$) 
on the $U(1)$ gauge field $A_\mu$ (which couples with the matter (Dirac) fields in a $U(1)$ gauge
invariant fashion). In these works [36,37], there is no presence of any  ambiguity parameter
$a$ and, therefore, there is no restriction of any kind. The interesting point is that,
even in this work, the expressions for the nilpotent
(anti-) dual-BRST charges are same as in (67). As a consequence,
the physicality condition on the harmonic (physical) state with these 
conserved and nilpotent charges is:
$Q_{(a)d}\; |\;phys> \;=\; 0 \Rightarrow E\; |\;phys> \;=\; 0, \dot E\; |\;phys> \;= \;0$ where
$E \sim \varepsilon^{\mu\nu} F_{\mu\nu}$ is the anomaly term in 2D.

It is to be emphasized that the consistency of the anomalous 2D Abelian 1-form gauge theory
is encoded in the physicality condition with the conserved and nilpotent (anti-) BRST and
(anti-) co-BRST charges. The anomalous terms, which are on the r.h.s. of the conservation
law $\partial_\mu J^\mu \sim  (a - 1) (\partial \cdot A) + \varepsilon^{\mu\nu} \partial_\mu A_\nu$
(see, e.g. [24],[25]), individually annihilate the harmonic (physical) state of the theory
due to $Q_{(a)b} |phys> = 0, Q_{(a)d} |phys> = 0$. Furthermore, these restrictions remain
invariant w.r.t. the time-evolution of the system because the physicality condition
implies that $(\partial \cdot A) |phys> = 0, \partial_0 (\partial \cdot A) |phys> = 0$
as well as $ (-\varepsilon^{\mu\nu} \partial_\mu A_\nu \equiv E) |phys> = 0, \dot E |phys> = 0$.

The precise reasons behind the similarity between the anomalous 2D Abelian 1-form gauge theory
and the free 4D Abelian 2-form gauge theory are {\it not} clear to us at the moment. This
issue is an interesting problem for our future investigations.
It would be nice to extend our present investigation to the 4D non-Abelian 2-form gauge
theory and establish its hidden connection with the anomalous 2D non-Abelian gauge
theory which has already been shown to be consistent and unitary [39]. To show that
the above theories and their possible modified versions are the field-theoretic models
for the Hodge theory, is a very challenging and demanding endeavor for us. We, at the moment, are
actively involved with the above-mentioned issues and we hope to report
about our results in our future publications [40].\\

\noindent
{\bf Acknowledgements}\\

\noindent
Financial support from the Department of Science and Technology, Government
of India, under the SERC project grant No: SR/S2/HEP-23/2006, is gratefully 
acknowledged. It is a great pleasure for us to thank the anonymous referee for some 
very useful suggestions.\\

\noindent
{\bf Appendix A: On (dual-) gauge transformations in 2-form theory}\\

\noindent
Let us begin with the following simple gauge-fixed Lagrangian density of 
the 4D Abelian 2-form gauge theory in the Feynman gauge 
\begin{eqnarray}
{\cal L}_0 = {\displaystyle \frac {1} {12} H_{\mu\nu\kappa} H^{\mu\nu\kappa} 
+ \frac{1}{2} (\partial^\nu B_{\nu\mu}) (\partial_\eta B^{\eta\mu})},
\end{eqnarray}
where the totally antisymmetric curvature tensor $ H_{\mu\nu\kappa} 
= \partial_\mu B _{\nu\kappa} + \partial_\nu B_{\kappa\mu} + \partial_\kappa
B_{\mu\nu} $ is derived from the 3-form $ H^{(3)} = d B^{(2)} = \frac {1} {3!}
(d x ^\mu \wedge d x^\nu \wedge d x^\kappa )$ $H_{\mu\nu\kappa}$. 
In the above, $ d = d x ^\mu \partial_\mu $ (with $ d^2 = 0 $) is the 
exterior derivative and the 2-form $ B^{(2)} = \frac{1}{2} 
(d x^\mu \wedge d x ^\nu) B_{\mu\nu} $ defines the antisymmetric 
$ (B_{\mu\nu} = - B_{\nu\mu})$ gauge potential $ B_{\mu\nu} $ of the present
gauge theory. In a similar fashion,
the gauge-fixing term is connected with the co-exterior derivative
$\delta = - * d * $
because $\delta B^{(2)} = - * d * B^{(2)} = (\partial_\nu B^{\nu\mu}) dx_\mu$. 
Here the ($*$) operation 
corresponds to the Hodge duality operation on the 4D Minkowski spacetime manifold.
The following
infinitesimal versions of the (dual-) gauge transformations
\begin{equation}
\delta_{dg} B_{\mu\nu} = - \varepsilon_{\mu\nu\eta\xi} \partial^\eta \Sigma^\xi, \qquad
\delta_g B_{\mu\nu} = - (\partial_\mu \alpha_\nu - \partial_\nu \alpha_\mu), 
\end{equation}
leave the gauge-fixing and the kinetic terms invariant, respectively, because
\begin{equation}
\delta_{dg} (\partial^\nu B_{\nu\mu}) = 0, \qquad \delta_g (H_{\mu\nu\eta}) = 0.
\end{equation}
This invariance is the reason behind the above nomenclature  
associated with the symmetry transformations. Thus,
the infinitesimal transformations  $\delta_{(d)g}$, in the above, correspond to
the (dual-) gauge transformations and $\Sigma_\mu$ and $\alpha_\mu$ are the corresponding
infinitesimal parameters.

It can be readily checked that the gauge-fixed Lagrangian density ${\cal L}_{0}$ 
transforms to a total spacetime derivative plus terms that are zero on the equation
of motion $\Box B_{\mu\nu} = 0$.
This statement can be captured by the following equations that represent the 
(dual-) gauge transformations, namely;
\begin{eqnarray}
\delta_{dg} {\cal L}_0 &=& - \frac{1}{2} \partial_\mu \Bigl [ H^{\mu\nu\eta} 
\varepsilon_{\nu\eta\xi\sigma} \partial^\xi \Sigma^\sigma - \varepsilon^{\mu\nu\eta\sigma}
(\partial^\xi H_{\xi\nu\eta}) \Sigma_\sigma \Bigr ] \nonumber\\
&-& \frac{1}{2} \varepsilon_{\xi\nu\eta\sigma} (\partial^\xi \Box B^{\nu\eta}) \Sigma^\sigma,
\end{eqnarray}
\begin{eqnarray}
\delta_{g} {\cal L}_0 &=& - \partial_\mu \Bigl [ (\partial^\nu B_{\nu\sigma}) (\partial^\mu \alpha^\sigma
- \partial^\sigma \alpha^\mu)  - \alpha^\sigma \partial^\mu (\partial^\nu B_{\nu\sigma})
+ \alpha^\sigma \partial_\sigma (\partial_\nu B^{\nu\mu}) \Bigr ] \nonumber\\
&-& (\partial^\nu \Box B_{\nu\mu}) \alpha^\mu. 
\end{eqnarray}
Thus, we note that the anomalous Abelian 1-form Lagrangian density 
(cf. Subsec. 3.1) and gauge-fixed version
of the Lagrangian density of the Abelian 2-form gauge theory have a similarity as far as
their properties under the (dual-) gauge transformations are concerned. We further point out
that the above observations are the prelude to the existence of the (anti-) BRST and the
(anti-) co-BRST symmetry transformations which we attempt below.\\

\noindent
{\bf Appendix B: On (anti-) BRST invariant  Lagrangian densities}\\

\noindent 
We begin here with the {\it basic} (anti-) BRST invariant Lagrangian densities
of the Abelian 2-form gauge theory in 4D [22] 
\begin{eqnarray}
&&{\cal L}_B = {\displaystyle \frac {1} {12} H_{\mu\nu\kappa} H^{\mu\nu\kappa} 
+ B^\mu (\partial^\nu B_{\nu\mu}) + \frac {1} {2} ( B \cdot B + \bar B \cdot \bar B) 
- \frac {1}{2} \partial _\mu \phi_1 \partial^\mu \phi_1}  \nonumber\\
&& + {\displaystyle (\partial_\mu \bar C_\nu - \partial_\nu \bar C_\mu ) 
(\partial^\mu C^\nu) + ( \partial \cdot C - \lambda) \rho 
+ (\partial \cdot \bar C + \rho) \lambda 
+ \partial_\mu \bar \beta \partial^\mu \beta},
\end{eqnarray}
\begin{eqnarray}
& {\cal L}_{\bar B} =  {\displaystyle \frac {1} {12} H_{\mu\nu\kappa} H^{\mu\nu\kappa} 
+ \bar B^\mu (\partial^\nu B_{\nu\mu}) + \frac {1} {2} (B \cdot B + \bar B \cdot \bar B)
-  \frac {1}{2} \partial _\mu \phi_1 \partial^\mu \phi_1} & \nonumber\\
& + (\partial_\mu \bar C_\nu - \partial_\nu \bar C_\mu) (\partial^\mu C^\nu)
+ (\partial \cdot C - \lambda) \rho + (\partial \cdot \bar C + \rho) \lambda 
+ \partial_\mu \bar \beta \partial^\mu \beta. &
\end{eqnarray} 
The above basic Lagrangian densities ${\cal L}_B $ and ${\cal L}_ {\bar B} $ are 
endowed with the gauge-fixing and Faddeev-Popov ghost terms as given 
below 
\begin{eqnarray}
& s_b s_{ab} \Big [2\beta \bar \beta + \bar C_\mu C^\mu - {\displaystyle \frac {1} {4}} B^{\mu\nu} 
B_{\mu\nu} \Big ] = B^\mu (\partial^\nu B_{\nu\mu}) + B \cdot \bar B 
+ \partial_\mu \bar \beta \partial^\mu \beta &\nonumber\\
& + (\partial_\mu \bar C_\nu - \partial_\nu \bar C_\mu )(\partial^\mu  C^\nu ) 
+ ( \partial \cdot C - \lambda) \rho 
+ (\partial \cdot \bar C + \rho) \lambda, &
\end{eqnarray} 
\begin{eqnarray}
& - s_{ab} s_b \Big [2\beta \bar \beta + \bar C_\mu C^\mu - {\displaystyle \frac {1} {4}} B^{\mu\nu} 
B_{\mu\nu} \Big ] = \bar B^\mu (\partial^\nu B_{\nu\mu}) + B \cdot \bar B 
+ \partial_\mu \bar \beta \partial^\mu \beta &\nonumber\\
& + (\partial_\mu \bar C_\nu - \partial_\nu \bar C_\mu )(\partial^\mu  C^\nu ) 
+ ( \partial \cdot C - \lambda) \rho 
+ (\partial \cdot \bar C + \rho) \lambda, & 
\end{eqnarray}     
where the nilpotent $(s_{(a)b}^2 = 0)$ and absolutely anticommuting
($s_b s_{ab} + s_{ab} s_b = 0$) (anti-) BRST symmetry transformations 
($ s_{(a)b} $) are [22]
\begin{eqnarray}
&& s_b B_{\mu\nu} = - (\partial_\mu C_\nu - \partial_\nu C_\mu ), \;\quad \; 
s_b C_\mu = -\partial_\mu \beta, \;\quad \; s_b \bar C_\mu = - B_\mu, \nonumber\\ 
&& s_b \phi_1 = \lambda, \quad s_b \bar \beta = - \rho, \quad 
s_b \bar B_\mu = - \partial_\mu \lambda,\; s_b[\rho, \lambda, B_\mu, \beta, \,
H_{\mu\nu\kappa}] = 0,
\end{eqnarray} 
\begin{eqnarray}
&& s_{ab} B_{\mu\nu} = - (\partial_\mu \bar C_\nu - \partial_\nu \bar C_\mu ), 
\quad s_{ab} \bar C_\mu = -\partial_\mu \bar \beta, 
\quad s_{ab} C_\mu = \bar B_\mu, \nonumber\\ 
&& s_{ab} \phi_1 = \rho, \quad s_{ab} \beta = - \lambda, \; 
s_{ab} B_\mu =  \partial_\mu \rho, \; 
s_{ab}[\rho, \lambda, \bar B_\mu, \bar \beta, H_{\mu\nu\kappa}] = 0.
\end{eqnarray}
We have obtained ${\cal L}_B $ and ${\cal L}_{\bar B} $ (cf. (85) and (86))
by exploiting $ B \cdot \bar B = \frac{1}{2} ( B \cdot B + \bar B \cdot \bar B ) 
- \frac{1}{2} \partial_\mu \phi_1 \partial^\mu \phi_1 $ because our present
theory is defined on a constrained submanifold described by the constrained field 
equation $ B_\mu - \bar B_\mu  =  \partial_\mu \phi_1$. Furthermore, the absolute
anticommutativity $(s_b s_{ab} + s_{ab} s_b = 0)$ of the above transformations is 
satisfied if and only if $ B_\mu - \bar B_\mu  =  \partial_\mu \phi_1$.
In particular, it can be checked that $ \{ s_b, s_{ab} \} B_{\mu\nu} \equiv 
( s_b s_{ab} + s_{ab} s_b ) B_{\mu\nu} = 0 $ is true only  
if the above equation is precisely respected\footnote{The (anti-) BRST 
invariant condition 
$B_\mu - \bar B_\mu - \partial_\mu \phi_1 = 0$ is obtained
in our earlier work on the geometrical superfield approach to free 4D Abelian 2-form gauge theory [20].
This condition is the analogue of the Curci-Ferrari resriction [21] that
appears in the context of the non-Abelian 1-form gauge theory. The above conditions
ensure the {\it absolute} anticommutativity of the off-shell nilpotent (anti-) BRST symmetries.}.

It can be checked that the Lagrangian densities (85) and (86) can be expressed 
in term of the above (anti-) BRST symmetry transformations as
\begin{eqnarray}
&& {\cal L}_B = {\displaystyle \frac {1} {12}} H_{\mu\nu\kappa} H^{\mu\nu\kappa}  +
s_b s_{ab} \Big [2\beta \bar \beta + \bar C_\mu C^\mu - \frac {1} {4} B^{\mu\nu} 
B_{\mu\nu} \Big ], \nonumber\\
&& {\cal L}_{\bar B} = {\displaystyle \frac {1} {12}} H_{\mu\nu\kappa} 
H^{\mu\nu\kappa} - s_{ab} s_b \Big [2\beta \bar \beta + \bar C_\mu C^\mu 
- \frac {1} {4} B^{\mu\nu} B_{\mu\nu} \Big ].
\end{eqnarray}
The BRST and anti-BRST invariance of (91) on the constrained surface 
$ B_\mu - \bar B_\mu - \partial_\mu \phi_1 = 0 $ becomes very clear and 
simple because of 

(i) the nilpotency $ (s_{(a)b}^2 = 0) $ and absolute anticommutativity $(s_b s_{ab} \;
 +  s_{ab} s_b = 0) $ of the (anti-) BRST symmetry transformations, and

(ii) the (anti-) BRST invariance of the curvature term (i.e. $ s_{(a)b} 
H_{\mu\nu\kappa} = 0) $.
The above statements can be corroborated by the following equations
\begin{eqnarray}
s_b {\cal L}_B &=& - \partial_\mu \Big[( \partial^\mu C^\nu - \partial^\nu C^\mu ) B_\nu 
+ \lambda B^\mu + \rho \partial^\mu \beta \Big ] \nonumber\\
&+& (\partial^\mu \lambda ) \Big [ B_\mu 
- \bar B_\mu - \partial_\mu \phi_1 \Big], 
\end{eqnarray}
\begin{eqnarray}
s_{ab} {\cal L}_{\bar B} &=& - \partial_\mu \Big[( \partial^\mu \bar C^\nu 
- \partial^\nu \bar C^\mu ) \bar B_\nu 
- \rho \bar B^\mu + \lambda \partial^\mu \bar \beta \Big ] \nonumber\\
&+& (\partial^\mu \rho ) \Big [ B_\mu 
- \bar B_\mu - \partial_\mu \phi_1 \Big ],
\end{eqnarray}
\begin{eqnarray}
& s_{ab} {\cal L}_B = - \partial_\mu 
\Big [(\partial^\mu \bar C^\nu- \partial^\nu \bar C^\mu ) B_\nu
- \rho \bar B^\mu  + \lambda \partial^\mu \bar \beta 
- \rho ( \partial_\nu B^{\nu\mu}) \Big ] & \nonumber\\
& + (\partial^\mu \bar C^\nu- \partial^\nu \bar C^\mu ) \partial_\mu
\Big [ B_\nu - \bar B_\nu - \partial_\nu \phi_1 \Big ] 
+ (\partial^\mu \rho ) \Big [B_\mu - \bar B_\mu - \partial_\mu \phi_1 \Big ],&
\end{eqnarray}
\begin{eqnarray}
& s_b {\cal L}_{\bar B} = - \partial_\mu \Big 
[(\partial^\mu C^\nu - \partial^\nu C^\mu ) \bar B_\nu  
+ \rho \partial^\mu \beta + \lambda B^\mu  + \lambda  
(\partial_\nu B^{\nu\mu})\Big ] &\nonumber\\
& - (\partial^\mu C^\nu - \partial^\nu C^\mu )  \partial_\mu 
\Big [ B_\nu - \bar B_\nu - \partial_\nu \phi_1 \Big] 
+ (\partial^\mu \lambda) \Big [B_\mu - \bar B_\mu - \partial_\mu \phi_1 \Big ].&
\end{eqnarray}
Thus, it is clear that if we impose the constraint field equation $(B_\mu - 
\bar B_\mu - \partial_\mu \phi_1 = 0)$, we shall have the following 
\footnote{These observations are important because we have seen a similar kind of 
symmetry structure in the case of anomalous 2D Abelian 1-form gauge theory in Sec. 3.}

(i) the absolute anticommutativity of the nilpotent (anti-) BRST 
symmetry transformations  $ s_{(a)b}$, and

(ii) the BRST and anti-BRST invariance of both the basic and equivalent 
Lagrangian densities ${\cal L}_B $ and ${\cal L}_{\bar B}$.

The constrained field equation $B_\mu - \bar B_\mu 
- \partial_\mu \phi_1 = 0$ is an (anti-) BRST invariant quantity 
(i.e. $ s_{(a)b} \big [B_\mu - \bar B_\mu - \partial_\mu \phi_1 \big] = 0$ ). 
As a side remark, it can be seen that the $(B \cdot \bar B)$, present in
(87) and (88), can also be expressed as 
\begin{eqnarray}
B \cdot \bar B = B \cdot B - B^\mu \partial_\mu \phi_1, \quad 
B \cdot \bar B = \bar B \cdot \bar B + \bar B^\mu \partial_\mu \phi_1.
\end{eqnarray}
In such a situation, the Euler-Lagrange equations of motion, that would emerge 
from ${\cal L}_B$ and  ${\cal L}_{\bar B}$, are [14,26] 
\begin{eqnarray}
B_\mu = - \frac {1} {2} (\partial^\nu B_{\nu\mu} - \partial_\mu \phi_1), \quad   
\bar B_\mu = - \frac {1} {2} (\partial^\nu B_{\nu\mu} + \partial_\mu \phi_1).    
\end{eqnarray}
The above expressions would lead to the constrained field equation $ B_\mu - 
\bar B_\mu - \partial_\mu \phi_1 = 0.$ Thus, we obtain the CF type 
constrained field equation $ (B_\mu - \bar B_\mu - \partial_\mu \phi_1 = 0)$ in  
two steps from the Lagrangian densities ${\cal L}_B$ and  ${\cal L}_{\bar B}$ by 
exploiting the equations of motion and subtracting one from the other. It would be,
however, very nice to obtain 

(i) the above constrained field equation (i.e. $ B_\mu - \bar B_\mu -
\partial_\mu \phi_1 = 0$) in one step by exploiting the equation of motion, and

(ii) the perfect (anti-) BRST symmetry invariance of the Lagrangian 
densities ${\cal L} _B$ and $ {\cal L}_{\bar B} $ without any imposition of 
$ B_\mu - \bar B_\mu - \partial_\mu \phi_1 = 0$.

To this end in mind, we add a Lagrange multiplier field ($ L^\mu $) in the Lagrangian 
densities in the following fashion (see, e.g. [22] for details).
\begin{eqnarray}
{\cal L}_{(L, B)} &=& \frac {1} {12} H_{\mu\nu\kappa} H^{\mu\nu\kappa} 
+ B^\mu (\partial^\nu B_{\nu\mu}) + \frac {1} {2} ( B \cdot B + \bar B \cdot \bar B) 
- \frac {1}{2} \partial _\mu \phi_1 \partial^\mu \phi_1 \nonumber\\
&+& \partial_\mu \bar \beta \partial^\mu \beta 
+ (\partial_\mu \bar C_\nu - \partial_\nu \bar C_\mu ) (\partial^\mu C^\nu)
+ ( \partial \cdot C - \lambda) \rho \nonumber\\ 
&+& (\partial \cdot \bar C + \rho) \lambda 
+ L^\mu (B_\mu - \bar B_\mu - \partial_\mu \phi_1),
\end{eqnarray}
\begin{eqnarray}
{\cal L}_{(L, \bar B)} &=& \frac {1} {12} H_{\mu\nu\kappa} H^{\mu\nu\kappa} 
+ \bar B^\mu (\partial^\nu B_{\nu\mu}) + \frac {1} {2} ( B \cdot B + \bar B \cdot \bar B) 
- \frac {1}{2} \partial _\mu \phi_1 \partial^\mu \phi_1 \nonumber\\
&+& \partial_\mu \bar \beta \partial^\mu \beta 
+ (\partial_\mu \bar C_\nu - \partial_\nu \bar C_\mu ) (\partial^\mu C^\nu)
+( \partial \cdot C - \lambda) \rho \nonumber\\
&+& (\partial \cdot \bar C + \rho) \lambda 
+ L^\mu (B_\mu - \bar B_\mu - \partial_\mu \phi_1).
\end{eqnarray}
It is straightforward to check that the above Lagrangian densities remain quasi-invariant
under the (anti-) BRST transformations if we take
\begin{eqnarray}
s_b L_\mu = - \partial_\mu \lambda, \qquad s_{ab} L_\mu = - \partial_\mu \rho.
\end{eqnarray}
The above transformations are consistent with the Euler-Lagrange equation of motion, 
nilpotency and absolute anticommutativity of the (anti-) BRST symmetry
transformations. Furthermore, it can be checked that, under the symmetry transformations 
(89) and (90), the above Lagrangian densities transform to the total spacetime derivatives 
(as given in (5) and (6)) without any imposition of the restriction 
like $ B_\mu - \bar B_\mu - \partial_\mu \phi_1 = 0 $. \\

\noindent
{\bf Appendix C: On (anti-) co-BRST invariant Lagrangian densities}\\

\noindent
The kinetic term $(\frac {1}{12} H^{\mu \nu \kappa} H_{\mu \nu \kappa})$ of the 
gauge field (cf. (85) and (86)) can be linearized by introducing a massless 
$(\Box \phi_2 = 0)$ scalar field $\phi_2$ and the auxiliary fields ${\cal B}_\mu$ 
and $\bar {\cal B}_\mu$. The ensuing equivalent Lagrangian densities 
\begin{eqnarray}
{\cal L}_{(B , \cal B)} &=& \frac {1}{2} \partial_\mu \phi_2 
\partial^\mu \phi_2- \frac {1}{2} {\cal B^\mu} \varepsilon _{\mu \nu \eta \kappa }
\partial^\nu B^{\eta \kappa}- \frac {1}{2} ({\cal B} \cdot {\cal B} 
+ \bar {\cal B} \cdot \bar {\cal B}) \nonumber\\
&+& B^\mu (\partial^\nu B_{\nu \mu})+ \frac {1}{2} (B \cdot B + \bar B \cdot \bar B)
- \frac {1}{2} \partial_\mu \phi_1 \partial^\mu \phi_1   
+ \partial_\mu \bar \beta \partial^\mu \beta \nonumber\\
&+& (\partial_\mu \bar C_\nu - \partial_\nu \bar C_\mu ) 
(\partial^\mu C^\nu )+(\partial \cdot C - \lambda ) \rho 
+ (\partial \cdot \bar C + \rho ) \lambda,
\end{eqnarray}
\begin{eqnarray}
{\cal L}_{(\bar B , \bar {\cal B})}  &=& \frac {1}{2} \partial_\mu \phi_2 
\partial^\mu \phi_2- \frac {1}{2} {\bar {\cal B}^\mu} \varepsilon _{\mu \nu \eta \kappa }
\partial^\nu B^{\eta \kappa}- \frac {1}{2} ({\cal B} \cdot {\cal B} 
+ \bar {\cal B} \cdot \bar {\cal B}) \nonumber\\
&+& \bar B^\mu (\partial^\nu B_{\nu \mu})+ \frac {1}{2} (B \cdot B + \bar B \cdot \bar B)
- \frac {1}{2} \partial_\mu \phi_1 \partial^\mu \phi_1   
+ \partial_\mu \bar \beta \partial^\mu \beta \nonumber\\
&+& (\partial_\mu \bar C_\nu - \partial_\nu \bar C_\mu ) 
(\partial^\mu C^\nu )+(\partial \cdot C - \lambda ) \rho 
+ (\partial \cdot \bar C + \rho ) \lambda,
\end{eqnarray}
respect, in addition to the (anti-) BRST symmetry transformations (89) and (90)\footnote{
In fact, in addition to the transformations (89) and (90), we also need the transformations
$s_{(a)b} \phi_2 = 0, s_{(a)b} {\cal B}_\mu = 0, s_{(a)b} \bar {\cal B}_\mu = 0$ for
the perfect symmetry invariance of the Lagrangian densities of our present 
theory (cf. (105)--(108) below).},
the following dual(co)-BRST and anti-dual(co)-BRST transformations
\begin{eqnarray} 
& s_d B_{\mu \nu} = - \varepsilon_{\mu \nu \eta \kappa} \partial^\eta \bar C^\kappa, 
\quad s_d \bar C_\mu = - \partial_\mu \bar \beta, \quad s_d C_\mu = - {\cal B}_\mu, 
\quad s_d \phi_2 = - \rho, & \nonumber \\
& s_d \beta = - \lambda, \; s_d \bar {\cal B}_\mu =  \partial_\mu \rho, 
\; s_d \big [\rho , \lambda, \bar \beta, \phi_1, 
{\cal B}_\mu, B_\mu, \bar B_\mu, (\partial^\nu B_{\nu \mu})\big ] = 0, & 
\end{eqnarray}
\begin{eqnarray}
& s_{ad} B_{\mu \nu} = - \varepsilon_{\mu \nu \eta \kappa} \partial^\eta C^\kappa, 
\quad s_{ad} C_\mu = \partial_\mu \beta, \quad s_{ad} \bar C_\mu = \bar {\cal B}_\mu, 
\quad s_{ad} \phi_2 = - \lambda, & \nonumber\\
& s_{ad} \bar \beta = \rho, \; s_{ad} {\cal B}_\mu = - \partial_\mu \lambda, 
\; s_{ad} \big [\rho , \lambda, \beta, \phi_1, 
\bar {\cal B}_\mu, \bar B_\mu,B_\mu, (\partial^\nu B_{\nu \mu})\big ] = 0. &
\end{eqnarray}
It is straightforward to check that the following are true, namely; 
\begin{eqnarray}
s_d {\cal L}_{(B , \cal B)} &=&  \partial_\mu  \Big [(\partial^\mu \bar C^\nu 
- \partial^\nu \bar C^\mu ) {\cal B}_\nu - \rho {\cal B}^\mu 
- \lambda \partial^\mu \bar \beta \Big ] \nonumber\\
&+& (\partial^\mu \rho ) \Big [{\cal B}_\mu - \bar {\cal B}_\mu - \partial_\mu \phi_2 \Big ],
\end{eqnarray}
\begin{eqnarray}
s_{ad} {\cal L}_{(\bar B , \bar {\cal B})} &=& \partial_\mu 
\Big [(\partial^\mu C^\nu- \partial^\nu C^\mu )\bar {\cal B}_\nu  
+ \rho \partial^\mu \beta + \lambda \bar {\cal B}^\mu \Big ] \nonumber\\
&+& (\partial^\mu \lambda) \Big [{\cal B}_\mu - \bar {\cal B}_\mu - \partial_\mu \phi_2 \Big ],
\end{eqnarray}
\begin{eqnarray}
&& s_d {\cal L}_{(\bar B , \bar {\cal B})} =  \partial_\mu \Big [(\partial^\mu 
\bar C^\nu - \partial^\nu \bar C^\mu ) {\cal B}_\nu - \frac {\rho}{2} 
\varepsilon^{\mu \nu \eta \kappa} \partial_\nu B_{\eta \kappa} 
- \rho {\cal B}^\mu - \lambda \partial^\mu \bar \beta \Big ] \nonumber\\
&& + ({\cal B}^\mu - \bar {\cal B}^\mu - \partial^\mu \phi_2 )(\partial_\mu \rho )
+ (\partial^\mu \bar C^\nu - \partial^\nu \bar C^\mu ) \partial_\mu 
\Big[{\cal B}_\nu - \bar {\cal B}_\nu - \partial_\nu \phi_2 \Big ],
\end{eqnarray}
\begin{eqnarray}
&& s_{ad} {\cal L}_{(B , \cal B)} = \partial_\mu \Big [(\partial^\mu C^\nu
- \partial^\nu C^\mu )\bar {\cal B}_\nu + \frac {\lambda}{2} 
\varepsilon^{\mu \nu \eta \kappa} \partial_\nu B_{\eta \kappa} 
+ \rho \partial^\mu \beta + \lambda \bar {\cal B}^\mu \Big ] \nonumber\\
&& + ({\cal B}^\mu - \bar {\cal B}^\mu - \partial^\mu \phi_2 )(\partial_\mu \lambda )
- (\partial^\mu C^\nu - \partial^\nu C^\mu ) \partial_\mu 
\Big[{\cal B}_\nu - \bar {\cal B}_\nu - \partial_\nu \phi_2 \Big ]. 
\end{eqnarray}
Thus, it is clear that, on the constraint surface 
$ {\cal B}_\mu - \bar {\cal B}_\mu - \partial_\mu \phi_2 = 0 $, the equivalent 
Lagrangian densities (101) and (102) are 
(anti-) co-BRST invariant. It is interesting to point out that similar kind of 
mathematical structure appears for the anomalous 2D gauge theory as well (see, Sec. 3).

Analogous to equation (91), the Lagrangian densities (101) and (102) can
be expressed as the sum of the full gauge-fixing term and (anti-) co-BRST exact
expressions as given below
\begin{eqnarray}
{\cal L}_{(B, {\cal B})} &=& \frac {1}{2} (B \cdot B + \bar B \cdot \bar B) 
+ B^\mu (\partial^\nu B_{\nu \mu})
- \frac {1}{2} \partial_\mu \phi_1 \partial^\mu \phi_1 \nonumber\\
&+& s_d s_{ad} \Big [2 \beta \bar \beta  + \bar C_\mu C^\mu - \frac {1} {4} B^{\mu\nu} 
B_{\mu\nu} \Big ], \nonumber\\
{\cal L}_{(\bar B, \bar {\cal B})} &=& \frac {1}{2} (B \cdot B + \bar B \cdot \bar B) 
+ \bar B^\mu (\partial^\nu B_{\nu \mu}) - \frac {1}{2} \partial_\mu \phi_1 
\partial^\mu \phi_1 \nonumber\\
&-& s_{ad} s_d \Big [2\beta \bar \beta  + \bar C_\mu C^\mu - \frac {1} {4} B^{\mu\nu} 
B_{\mu\nu}\Big ].
\end{eqnarray}
In this form, the (anti-) co-BRST invariance of the Lagrangian densities (101) and 
(102) becomes very simple (on the constrained submanifold defined by the field 
equation $ {\cal B}_\mu - \bar {\cal B}_\mu - \partial_\mu \phi_2 = 0 $) because of 

(i) the nilpotency of the (anti-) co-BRST symmetry transformations, and 

(ii) the invariance of the total gauge-fixing term 
\big (i.e. $ s_{(a)d} \; \big [\frac {1}{2} (B \cdot B + \bar B \cdot \bar B)
+ \bar B^\mu (\partial^\nu B_{\nu \mu}) - \frac {1}{2} \partial_\mu \phi_1 
\partial^\mu \phi_1 \big ] = 0, s_{(a)d} \; \big [\frac {1}{2} (B \cdot B + \bar B \cdot \bar B)
+  B^\mu (\partial^\nu B_{\nu \mu}) - \frac {1}{2} \partial_\mu \phi_1 
\partial^\mu \phi_1 \big ] = 0$ \big ) under the (anti-) co-BRST 
transformations $ s_{(a)d} $.

The explicit expression, modulo some total spacetime derivative terms, 
for the following combination, namely;
\begin{eqnarray}
& s_d s_{ad} \Big [ 2\beta \bar \beta  + \bar C_\mu C^\mu - \frac {1} {4} B^{\mu\nu} 
B_{\mu\nu}\Big ] = (\partial_\mu \bar C_\nu - \partial_\nu \bar C_\mu ) 
(\partial^\mu C^\nu )& \nonumber\\ &+ (\partial \cdot C - \lambda ) \rho  
 + (\partial \cdot \bar C + \rho ) \lambda 
+ \partial_\mu \bar \beta \partial^\mu \beta - {\cal B} \cdot \bar {\cal B}
- \frac {1}{2} {\cal B}^\mu  \varepsilon_{\mu\nu\eta\kappa} \partial^\nu 
B^{\eta\kappa},&
\end{eqnarray}
leads to the derivation of the Lagrangian density (101). In a similar fashion,
the following relationship (modulo some total spacetime derivative terms):
\begin{eqnarray}
&- s_{ad} s_d \Big [2\beta \bar \beta  + \bar C_\mu C^\mu - \frac {1} {4} B^{\mu\nu} 
B_{\mu\nu} \Big ] = (\partial_\mu \bar C_\nu - \partial_\nu \bar C_\mu ) 
(\partial^\mu C^\nu ) & \nonumber\\
&+(\partial \cdot C - \lambda ) \rho + (\partial \cdot \bar C + \rho ) \lambda 
+ \partial_\mu \bar \beta \partial^\mu \beta - {\cal B} \cdot \bar {\cal B}
- \frac {1}{2} \bar {\cal B}^\mu  \varepsilon_{\mu\nu\eta\kappa} \partial^\nu 
B^{\eta\kappa},&
\end{eqnarray}
leads to the derivation of (102) if we use 
\begin{eqnarray}
{\cal B} \cdot \bar {\cal B} = \frac {1}{2} ({\cal B} \cdot {\cal B}
+ \bar {\cal B} \cdot \bar {\cal B}) - \frac {1}{2} \partial_\mu \phi_2 \partial^\mu \phi_2,
\end{eqnarray}
that emerges due to the CF type of restriction ${\cal B}_\mu - \bar {\cal B}_\mu 
- \partial_\mu \phi_2 = 0. $ It is worthwhile, once again, to point out    
that the Lagrangian densities (101) and (102) are equivalent (on the constraint 
submanifold defined by the field equation  ${\cal B}_\mu - \bar {\cal B}_\mu 
- \partial_\mu \phi_2 = 0 $) and both of them respect the (anti-) co-BRST 
symmetry transformations. Analogous to (98) and (99), we can also write (101) and (102)
by incorporating $M^\mu ({\cal B}_\mu - \bar {\cal B}_\mu - \partial_\mu \phi_2)$
(with $M_\mu$ as Lagrange multiplier field)
which would respect the (anti-) co-BRST symmetries without any imposition. 
These issues have been taken into account in (1) and (2).

\end{document}